\documentclass{aa}  

\usepackage{graphicx}
\usepackage{txfonts}
\usepackage{amssymb}
\usepackage{lineno}
\usepackage{txfonts}
\usepackage{natbib}   
\usepackage{amstext}
\usepackage{mathabx}
\usepackage{multirow}
\usepackage{ulem}
\usepackage{color}
\usepackage{pifont}
\usepackage{lscape}

\usepackage{hyperref}
\newcommand{\science}{Science }

\newcommand{\kms}{\,km$\cdot$s$^{-1}$}

\newcommand{\Fluxunit}{\,cm$^{-2}\cdot$s$^{-1}$}
\newcommand{\Kunit}{\,cm$^{2}\cdot$s$^{-1}$}

\newcommand{\dix}[1]{$\times10^{#1}$}
\newcommand{\fig}[1]{Fig.~\ref{#1}}

\newcommand{\tab}[1]{Table~\ref{#1}}

\newcommand{\Result}[3]{({#1}$\pm${#2})$\times$10$^{#3}$}

\newcommand{\degre}{\ensuremath{^\circ}}

\begin{document} 


   \title{{\it Herschel}\thanks{{\it Herschel} is an ESA space observatory with science instruments provided by European-led Principal Investigator consortia and with important participation from NASA.} map of Saturn's stratospheric water, delivered by the plumes of Enceladus}


   \author{T.~Cavali\'e\inst{1}\inst{2}
          \and
          V.~Hue\inst{3}
          \and
          P.~Hartogh\inst{4}
          \and
          R.~Moreno\inst{2}
          \and
          E.~Lellouch\inst{2}
          \and
          H.~Feuchtgruber\inst{5}
          \and
          C.~Jarchow\inst{4}
          \and
          T.~Cassidy\inst{6}
          \and
          L.N.~Fletcher\inst{7}
          \and
          F.~Billebaud\inst{1}
          \and
          M.~Dobrijevic\inst{1}
          \and
          L.~Rezac\inst{4}
          \and
          G.S.~Orton\inst{8}
          \and
          M.~Rengel\inst{4}
          \and
          T.~Fouchet\inst{2}
          \and
          S.~Guerlet\inst{9}
           }

   \institute{Laboratoire d'Astrophysique de Bordeaux, Univ. Bordeaux, CNRS, B18N, all\'ee Geoffroy Saint-Hilaire, 33615 Pessac, France\\
              \email{thibault.cavalie@u-bordeaux.fr}
         \and
             LESIA, Observatoire de Paris, PSL Research University, CNRS, Sorbonne Universit\'es, UPMC Univ. Paris 06, Univ. Paris Diderot, Sorbonne Paris Cit\'e, F-92195 Meudon, France\\
          \and
             Southwest Research Institute, San Antonio, TX 78228, United States\\
          \and
             Max-Planck-Institut f\"ur Sonnensystemforschung, 37077 G\"ottingen, Germany\\
          \and
             Max Planck Institut f\"ur Extraterrestrische Physik, Garching, Germany\\
          \and
             Laboratory for Atmospheric and Space Physics, University of Colorado, Boulder, CO 80303, USA\\
          \and
             Department of Physics \& Astronomy, University of Leicester, University Road, Leicester LE1 7RH, UK\\
          \and
             Jet Propulsion Laboratory, California Institute of Technology, 4800 Oak Grove Drive, Pasadena, CA 91109, USA\\
          \and
             Sorbonne Universit\'es, UPMC Paris 06, UMR 8539, LMD, F-75005 Paris, France\\
             }

   \date{Received May 24, 2019; Accepted July 23, 2019}

 
  \abstract
   {The origin of water in the stratospheres of Giant Planets has been an outstanding question ever since its first detection by ISO some 20\, years ago. Water can originate from interplanetary dust particles, icy rings and satellites and large comet impacts. Analysis of \textit{Herschel} Space Observatory observations have proven that the bulk of Jupiter's stratospheric water was delivered by the Shoemaker-Levy~9 impacts in 1994. In 2006, the Cassini mission detected water plumes at the South Pole of Enceladus, placing the moon as a serious candidate for Saturn's stratospheric water. Further evidence was found in 2011, when \textit{Herschel} demonstrated the presence of a water torus at the orbital distance of Enceladus, fed by the moon's plumes. Finally, water falling from the rings onto Saturn's uppermost atmospheric layers at low latitudes was detected during the final orbits of Cassini's end-of-mission plunge into the atmosphere. }
   {In this paper, we use \textit{Herschel} mapping observations of water in Saturn's stratosphere to identify its source. }
   {Several empirical models are tested against the \textit{Herschel}-HIFI  and -PACS observations, which were collected on December 30, 2010, and January 2$^\mathrm{nd}$, 2011 (respectively). }
   {We demonstrate that Saturn's stratospheric water is not uniformly mixed as a function of latitude, but peaking at the equator and decreasing poleward with a Gaussian distribution. We obtain our best fit with an equatorial mole fraction 1.1\,ppb and a half-width at half-maximum of 25\degre, when accounting for a temperature increase in the two warm stratospheric vortices produced by Saturn's Great Storm of 2010-2011. }
   {This work demonstrates that Enceladus is the main source of Saturn's stratospheric water.}

   \keywords{Planets and satellites: individual: Saturn -- 
   Planets and satellites: atmospheres -- 
   Planets and satellites: individual: Enceladus}

   \maketitle

\section{Introduction}
The interiors of Giant Planets are supposedly rich in oxygen \citep{Owen2003,Gautier2001,Hersant2004}. In the deep hot layers, where thermochemical equilibrium prevails, H$_2$O is the most abundant oxygen species. Oxygen species are transported from deep down towards higher levels, but only  CO (and CO$_2$ in Jupiter and Saturn) can reach the stratosphere because of the tropopause cold trap \citep{Lodders1994,Wang2015,Cavalie2017}. H$_2$O was thus only observed in the tropospheric saturation layers of Jupiter and Saturn \citep{Larson1975,deGraauw1997}. The detection of oxygen species (H$_2$O, CO and CO$_2$) in the stratospheres of the Giant Planets and Titan \citep{Feuchtgruber1997,Bezard2002,Lellouch2002,Coustenis1998,Samuelson1983,Noll1986,Marten1993,Burgdorf2006} thus demonstrated that the upper atmospheres of the giant planets are contaminated by external sources from their close and/or more distant environments.

While oxygen-rich interplanetary dust particles (IDP) produced from asteroid collisions and from comet activity are a ubiquitous source \citep{Landgraf2002,Poppe2016}, it seems to be the main H$_2$O source only at Uranus and Neptune \citep{Moses2017} and the emerging overall picture however looks more complex. Other sources can actually be at work, like local sources from planetary icy environments (rings, satellites) \citep{Strobel1979,Connerney1986,Prange2006,Waite2018,Mitchell2018,Hsu2018}, or cometary ``Shoemaker-Levy 9 (SL9) type'' impacts \citep{Lellouch1995}. In Jupiter's stratosphere, H$_2$O, CO, and CO$_2$ come from the SL9 comet fragments \citep{Bezard2002,Lellouch2002,Lellouch2006,Cavalie2008c,Cavalie2012,Cavalie2013}. An older comet-impact component was even proposed to explain its the CO in the lower stratosphere \citep{Bezard2002}, even if IDP are another option \citep{Moses2017}. Comets are also the probable source of CO beyond Jupiter (see review in \citealt{Mandt2015} and \citealt{Moses2017}), as seen in Saturn \citep{Cavalie2009,Cavalie2010}, Uranus \citep{Cavalie2014}, and Neptune \citep{Lellouch2005,Lellouch2010,Luszcz-Cook2013,Moreno2017}. 

In Saturn, neither ISO nor SWAS disk-averaged observations \citep{Feuchtgruber1997,Bergin2000} had sufficient spectral resolution and/or high enough signal-to-noise ratio to identify the source of external H$_2$O. Now, a comet impact is probably not the cause of the observed stratospheric H$_2$O \citep{Moses2017} because of the contradiction between (i) the relatively ancient impacts required to fit the CO observations of \citet{Cavalie2010} and (ii) the relatively short diffusion timescale from the deposition level of cometary material in such impacts \citep{Lellouch1995,Moreno2003} down to the H$_2$O condensation level \citep{Ollivier2000,Moses2000,Moses2005}, which is located between the 1 and a few mbar level, depending on the latitude. In addition, the H$_2$O/CO ratio in Saturn's stratosphere ($\sim$0.15 -- \citealt{Moses2017}) is too large to be characteristic of a cometary impact, that delivers mostly CO  (H$_2$O/CO$<$0.01 in Neptune). This tends to indicate a source that is steadier than a discrete comet impact. Previous observations of Saturn with \textit{Herschel}-HIFI led to the detection of an H$_2$O torus at the orbit of Enceladus \citep{Hartogh2011}, fed by the plumes of this moon \citep{Hansen2006,Porco2006,Waite2006}. The fate of H$_2$O from this torus is eventually to spread in Saturn's system \citep{Cassidy2010} and a fraction is predicted to fall into Saturn's stratosphere with a distribution centered around the equator. Based on these models, and comparing with the Herschel observations, \citet{Hartogh2011} tentatively concluded that Enceladus was the source of Saturn's stratospheric water. It should be noted that a fraction is also expected to feed Titan's atmosphere as well, and models show that the flux expected at Titan and originating from the Enceladus water torus could explain the observations \citep{Moreno2012,Lara2014,Dobrijevic2014,Rengel2014,Hickson2014}. 

Recent \textit{in situ} measurements of Cassini during the proximal orbits of its end-of-mission demonstrated the existence of a flux of material from the rings to Saturn's atmosphere. The Magnetospheric Imaging Instrument (MIMI), the Ion and Neutral Mass Spectrometer (INMS), and the Cosmic Dust Analyzer (CDA) instruments have measured an infall of material originating from the D-ring: (i) neutral icy grains within 2\degre~around the equator, (ii) various gases including H$_2$O (concentrated at the 24\% level) in a latitudinal band of 8\degre~centered on the equator \citep{Waite2018}, (iii) and charged grains transported to higher latitudes along the magnetic field lines \citep{Mitchell2018,Hsu2018}. This latter phenomenon is the long anticipated ring rain phenomenon \citep{Connerney1984,Connerney1986,Prange2006,Moore2010,Moore2015,ODonoghue2017}. The mass flux of infalling dust and gas are estimated to be $\sim$5\,kg/s \citep{Mitchell2018} and $\sim$$10^4$\,kg/s \citep{Waite2018,Perry2018}, respectively. The sources that can explain the presence of H$_2$O in Saturn's stratosphere are thus in principle IDP, Enceladus plumes and Saturn's rings. 

Disentangling the various sources of externally supplied water in Giant Planet stratospheres, and thus in Saturn, was a key objective of the \textit{Herschel} key program HssO (\textit{Herschel} Solar System Observations) \citep{Hartogh2009}. In section~\ref{Observations} of this paper, we present the first disk-resolved mapping observations of H$_2$O in Saturn's stratosphere and a disk-averaged observation, obtained with the Photodetector Array Camera and Spectrometer (PACS) \citep{Poglitsch2010} and the Heterodyne Instrument for the Far Infrared (HIFI) \citep{deGraauw2010} (respectively), two instruments onboard the ESA \textit{Herschel} Space Observatory \citep{Pilbratt2010}. Using a combination of models presented in section~\ref{Models}, we derive the H$_2$O meridional distribution in Saturn's stratosphere in section~\ref{Results}. We discuss its origin according to our results in section~\ref{Discussion} and give our conclusion in section~\ref{Conclusion}.

\section{Observations \label{Observations}}
In the following sections, we present a summary of the performed observations (\tab{Obs_list}) and the relevant geometry of Saturn (as obtained from the JPL/Horizons database)

\begin{table*}
  \caption{Summary of the \textit{Herschel}-PACS and \textit{Herschel}-HIFI observations of Saturn and the relevant observation geometry. }             
  \label{Obs_list}      
  \begin{center}          
  \begin{tabular}{ll||ll}
    \hline
    Start date & 2010-12-31 03:33:17 & Date for computations & 2010-12-31 03:40:00 \\ 
    OD & 596 & Ang. Diam. & 17.20\arcsec \\
    Obs. ID & 1342212192 &$\lambda_\mathrm{obs}$ & 226.63\degre \\
    $\Delta t$ [s] & 744 & $\phi_\mathrm{obs}$ & 12.42\degre \\
    Instrument & HIFI &  $\lambda_\Sun$ & 231.99\degre \\
    Frequency & 1097.365\,GHz & $\phi_\Sun$ & 9.28\degre \\
    & & $a_{NP}$ & 357.29\degre \\
    \hline   
    Start date & 2011-01-02 10:32:21 & Date for computations & 2011-01-02 10:47:00 \\
    OD & 598 & Ang. Diam. & 17.26\arcsec \\
    Obs. ID & 1342212275 &$\lambda_\mathrm{obs}$ & 288.72\degre \\
    $\Delta t$ [s] & 1693 & $\phi_\mathrm{obs}$ & 12.46\degre \\
    Instrument & PACS &  $\lambda_\Sun$ & 294.11\degre \\
    Wavelength & 66.44\,$\mu$m & $\phi_\Sun$ & 9.32\degre \\
    & & $a_{NP}$ & 357.30\degre \\
    \hline   
  \end{tabular}
  \end{center}
  \small{\underline{Note:} OD means Herschel operational day, $\Delta t$ is the total integration time. Ang. Diam. is Saturn's equatorial angular diameter, $\lambda_\mathrm{obs}$ and $\phi_\mathrm{obs}$ are the longitude and latitude (respectively) of the sub-observer point, $\lambda_\Sun$ and $\phi_\Sun$ are the longitude and  latitude (respectively) of the sub-solar point, $a_{NP}$ is the North Polar angle. The solar longitude of Saturn ($L_S$) is 17\degre. All latitudes in this table are planetographic, and all longitudes are System III West longitudes. The physical parameters of Saturn (right column) have been obtained from the JPL/Horizons database.
}
\end{table*}

   \subsection{PACS observations}
   We observed Saturn with \textit{Herschel}/PACS in the framework of the HssO Key Program on January 2, 2011, with the aim of mapping the distribution of H$_2$O in the stratosphere of the planet. We used PACS in its line spectrometry mode \citep{Poglitsch2010}. The integral field spectrometer consists of 5$\times$5 spatial pixels ($\sim$50\arcsec$\times$50\arcsec~on sky), each covering a short instantaneous wavelength range sampled by 16 spectral pixels. Our observation of Saturn consists of a 3$\times$3 raster map with 3\arcsec~separation at 66.44\,$\mu$m. We thus pointed the 25-pixel array over 9 different positions to record an oversampled 225-point map of Saturn at 66.44\,$\mu$m. For the observation of such an intense far-infrared continuum source, we had to use a non-standard mode, in which the spectrometer readout electronics were configured to the shortest possible reset intervals of 1/32\,s, to avoid detector saturation. The half-power beam width (HPBW) of \textit{Herschel} is 9.4\arcsec~for the PACS observations, which results in covering latitudes from $-23$\degre~to 45\degre~for a beam centered on the planet. The spectral resolving power is $\sim$2500-3000. 

\begin{landscape}
\begin{figure}[t]
\begin{center}
   \includegraphics[width=7.5cm,keepaspectratio]{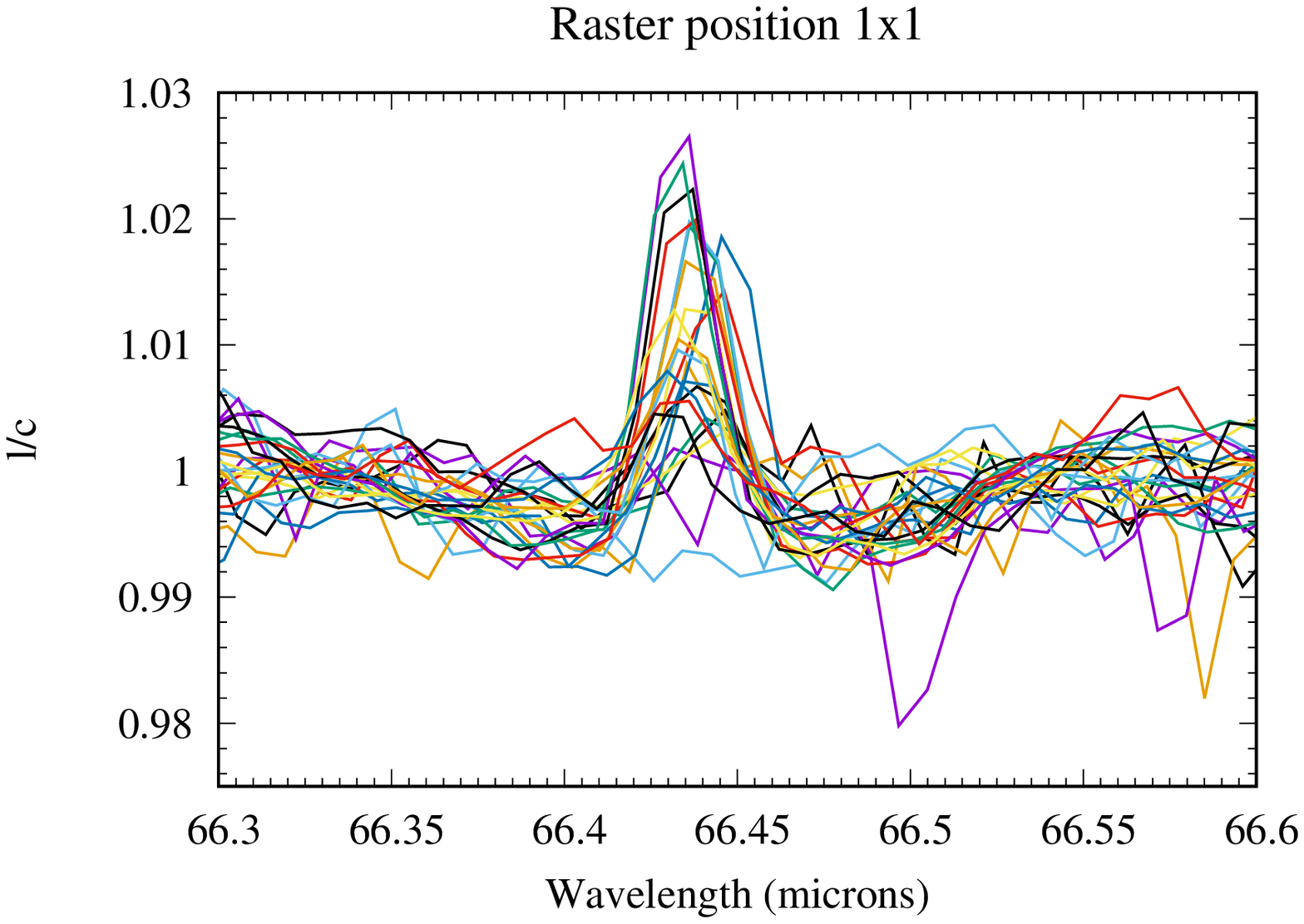}
   \includegraphics[width=7.5cm,keepaspectratio]{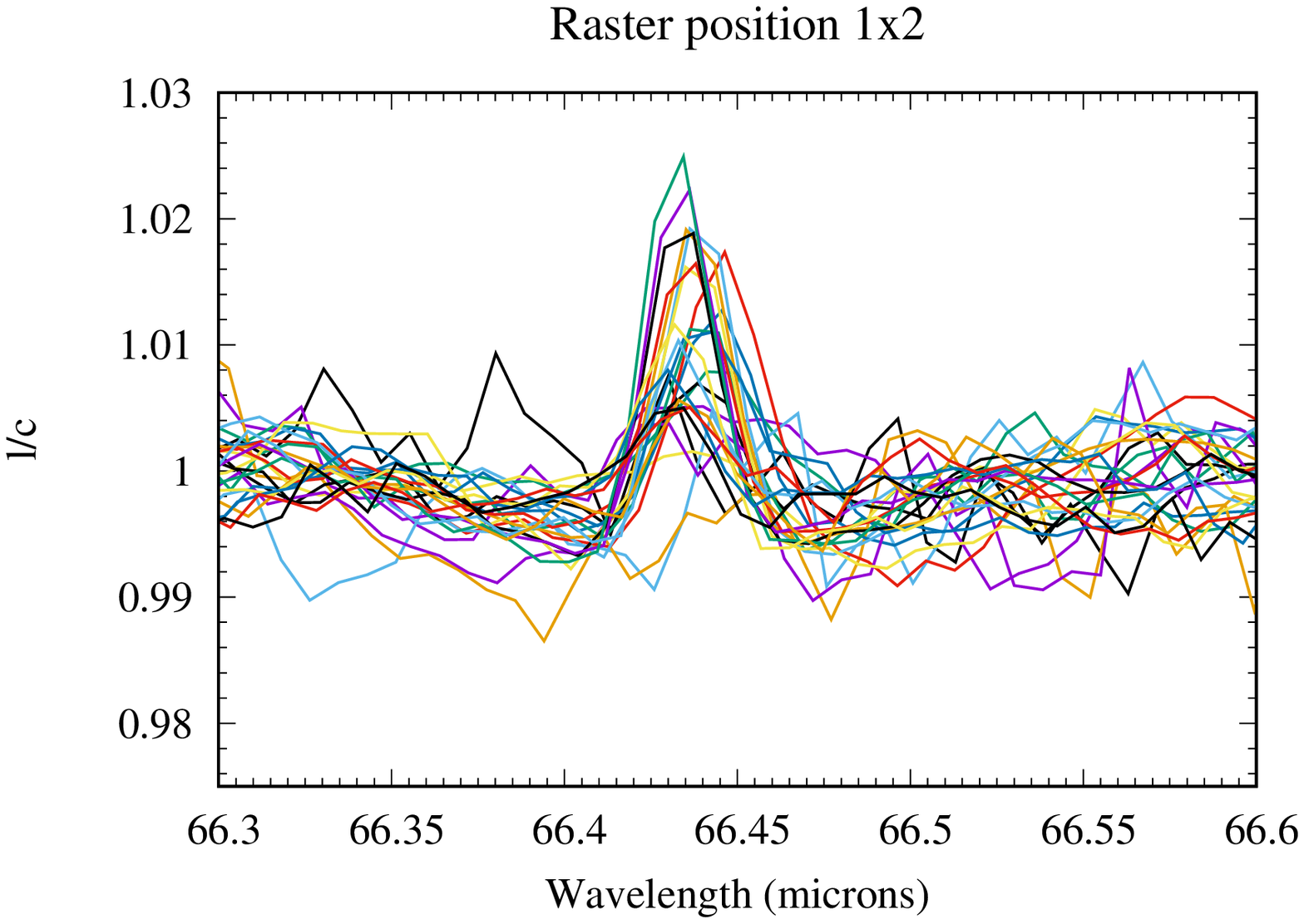}
   \includegraphics[width=7.5cm,keepaspectratio]{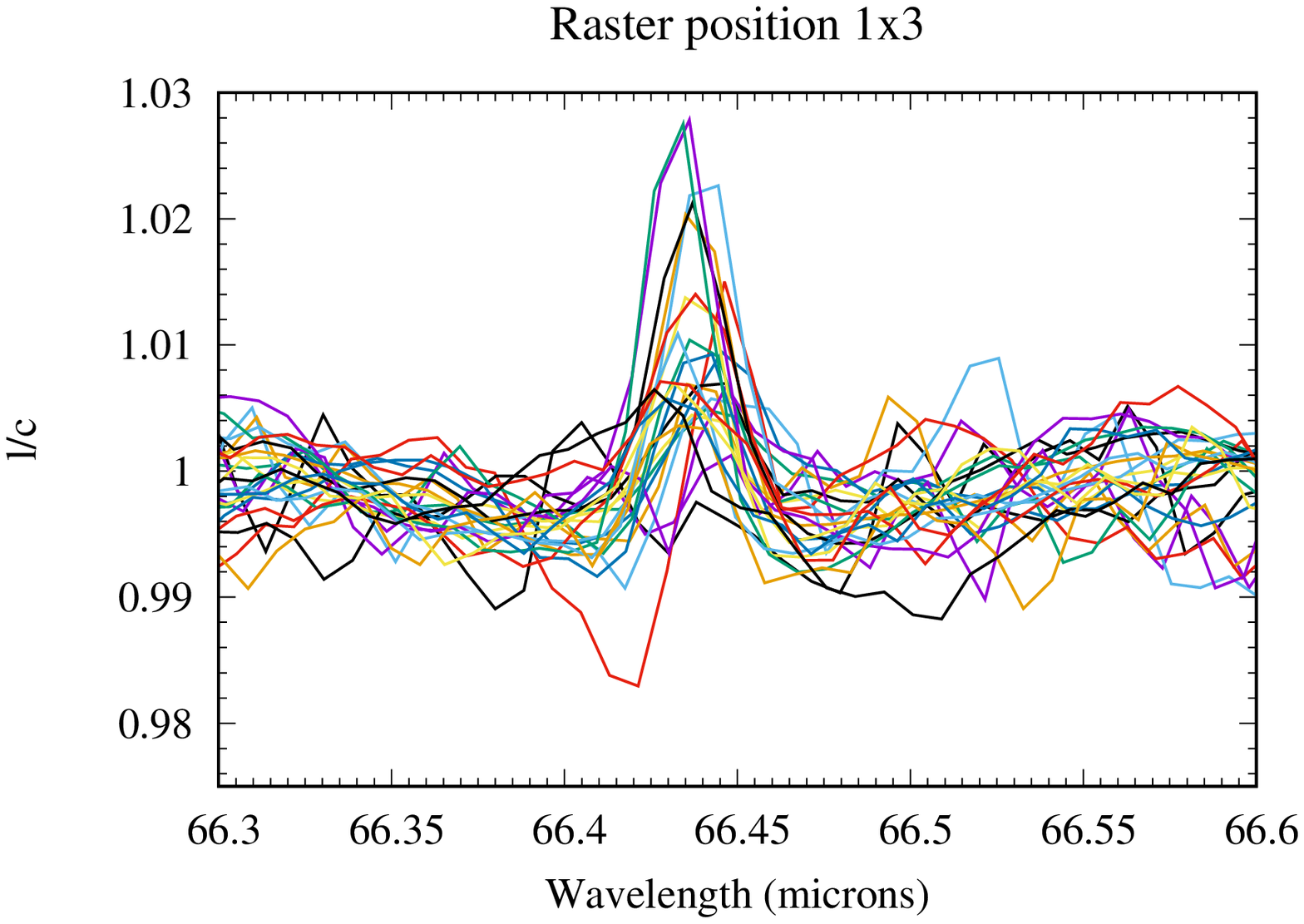}\\
   \includegraphics[width=7.5cm,keepaspectratio]{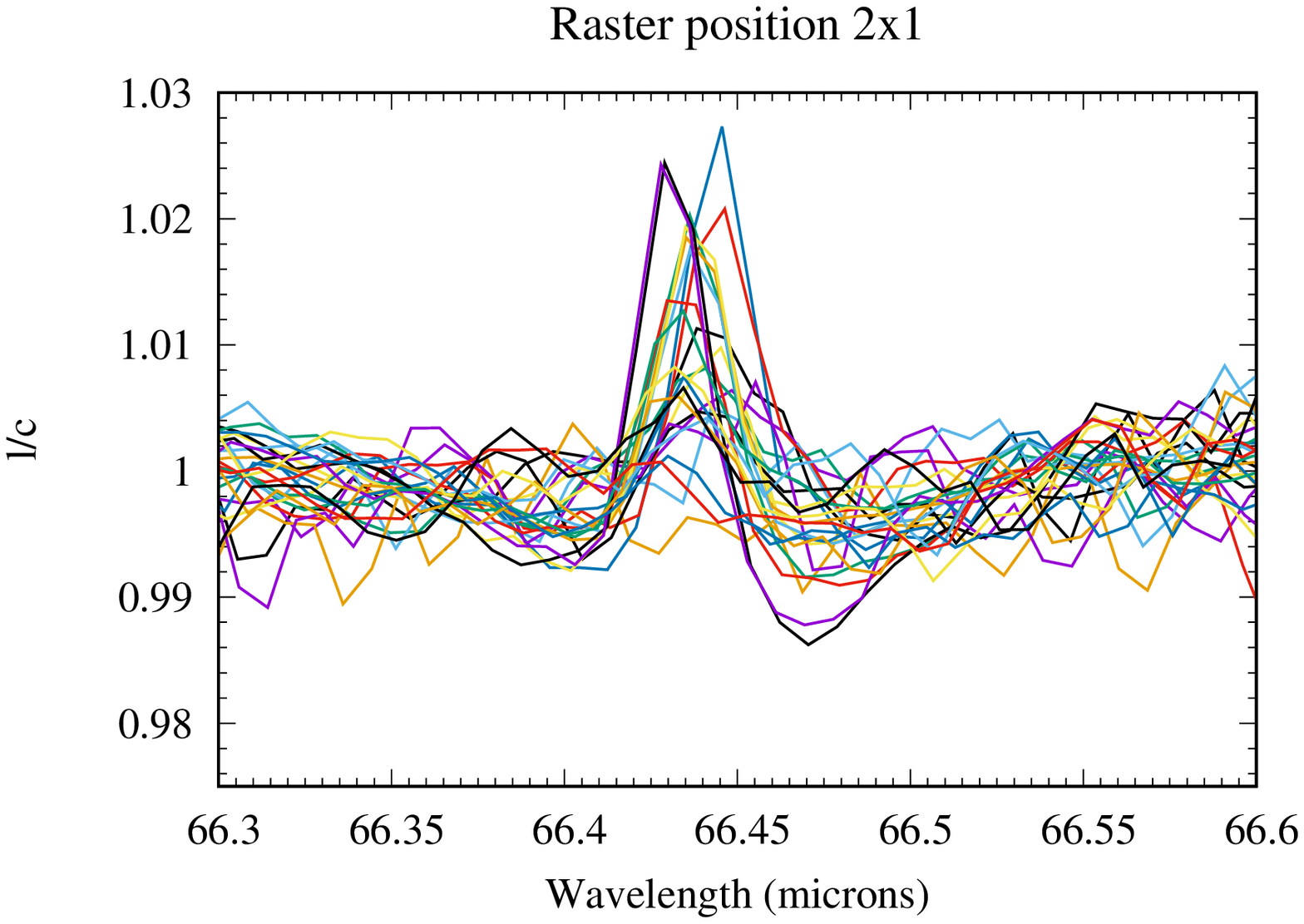}
   \includegraphics[width=7.5cm,keepaspectratio]{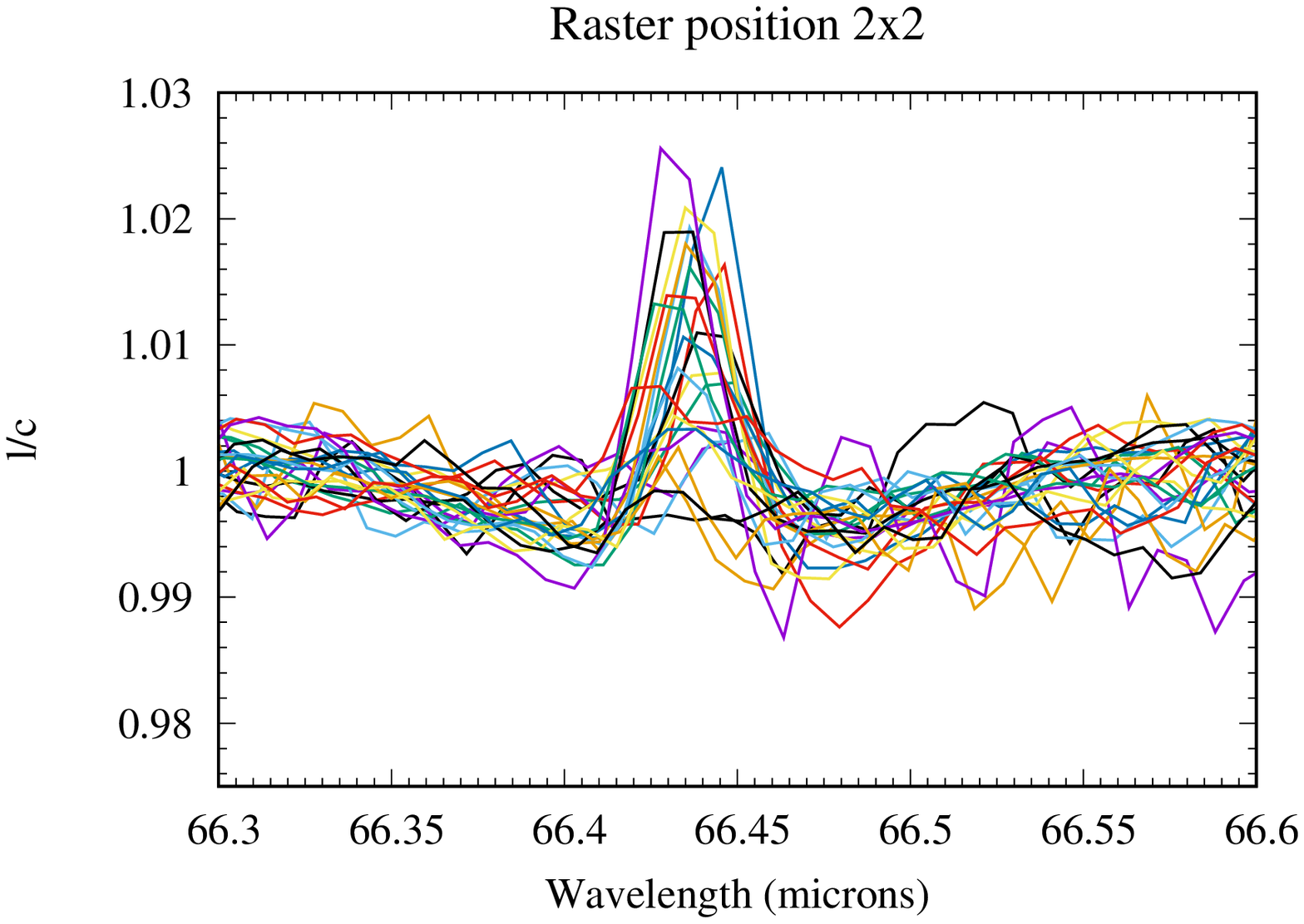}
   \includegraphics[width=7.5cm,keepaspectratio]{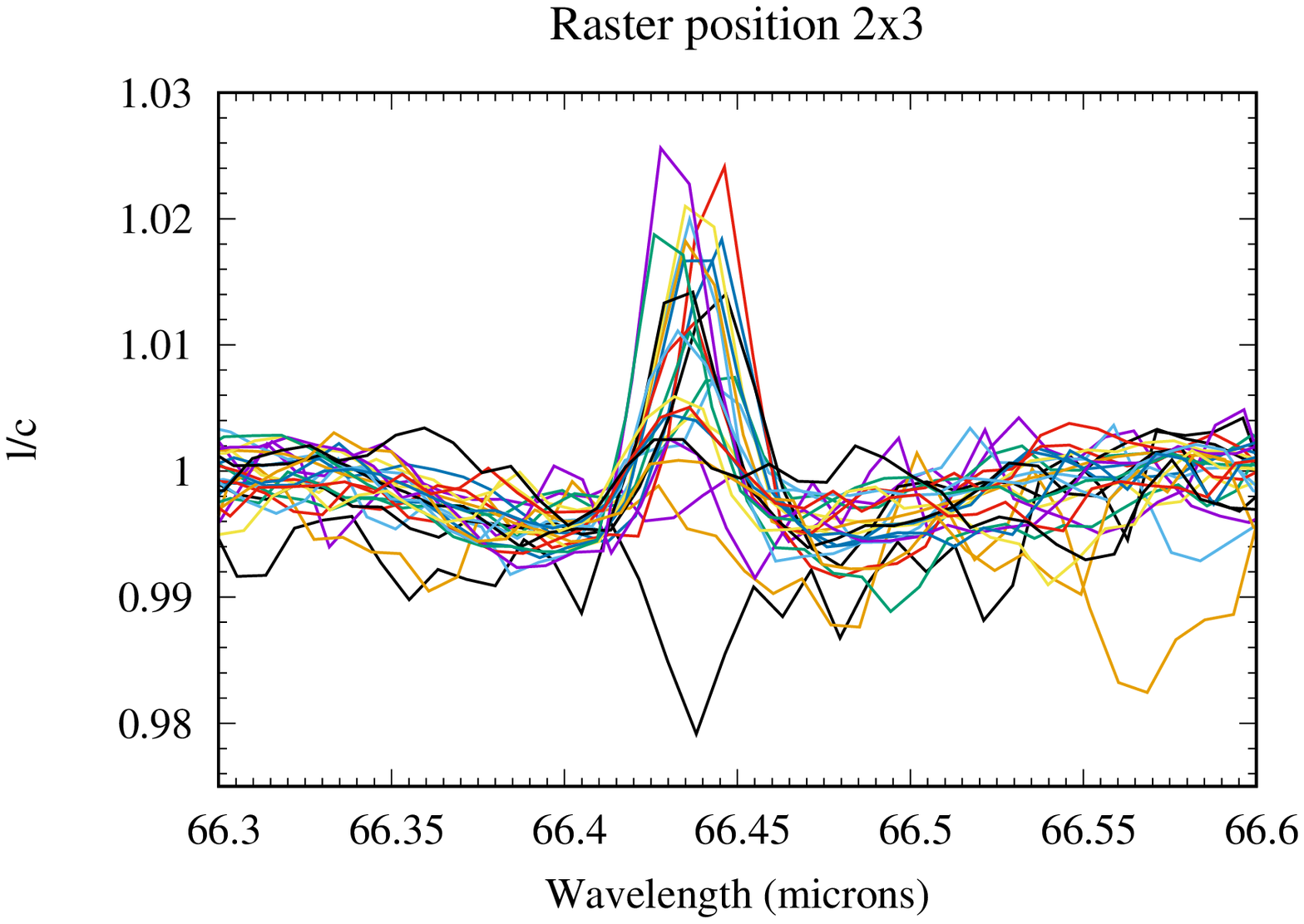}\\
   \includegraphics[width=7.5cm,keepaspectratio]{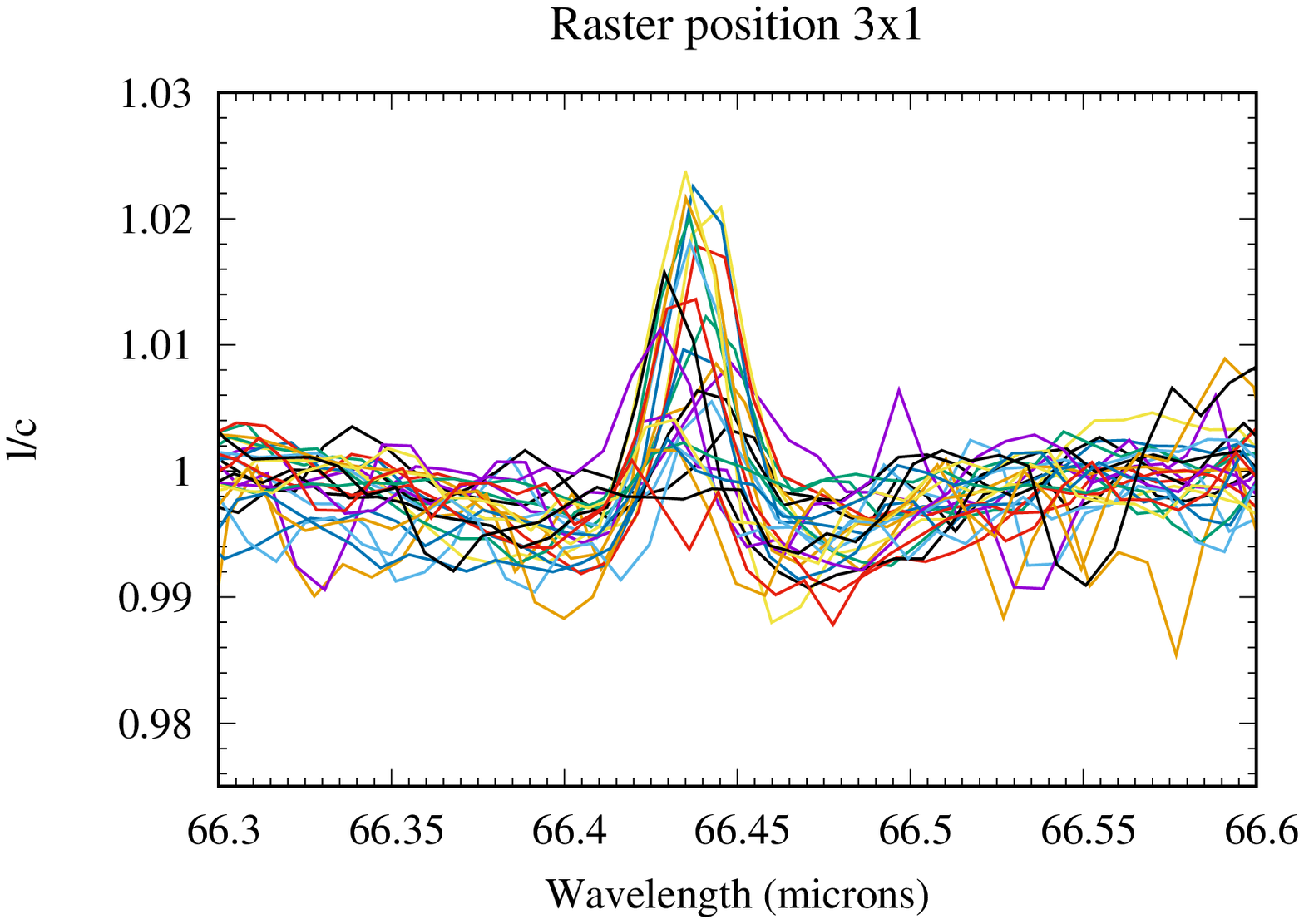}
   \includegraphics[width=7.5cm,keepaspectratio]{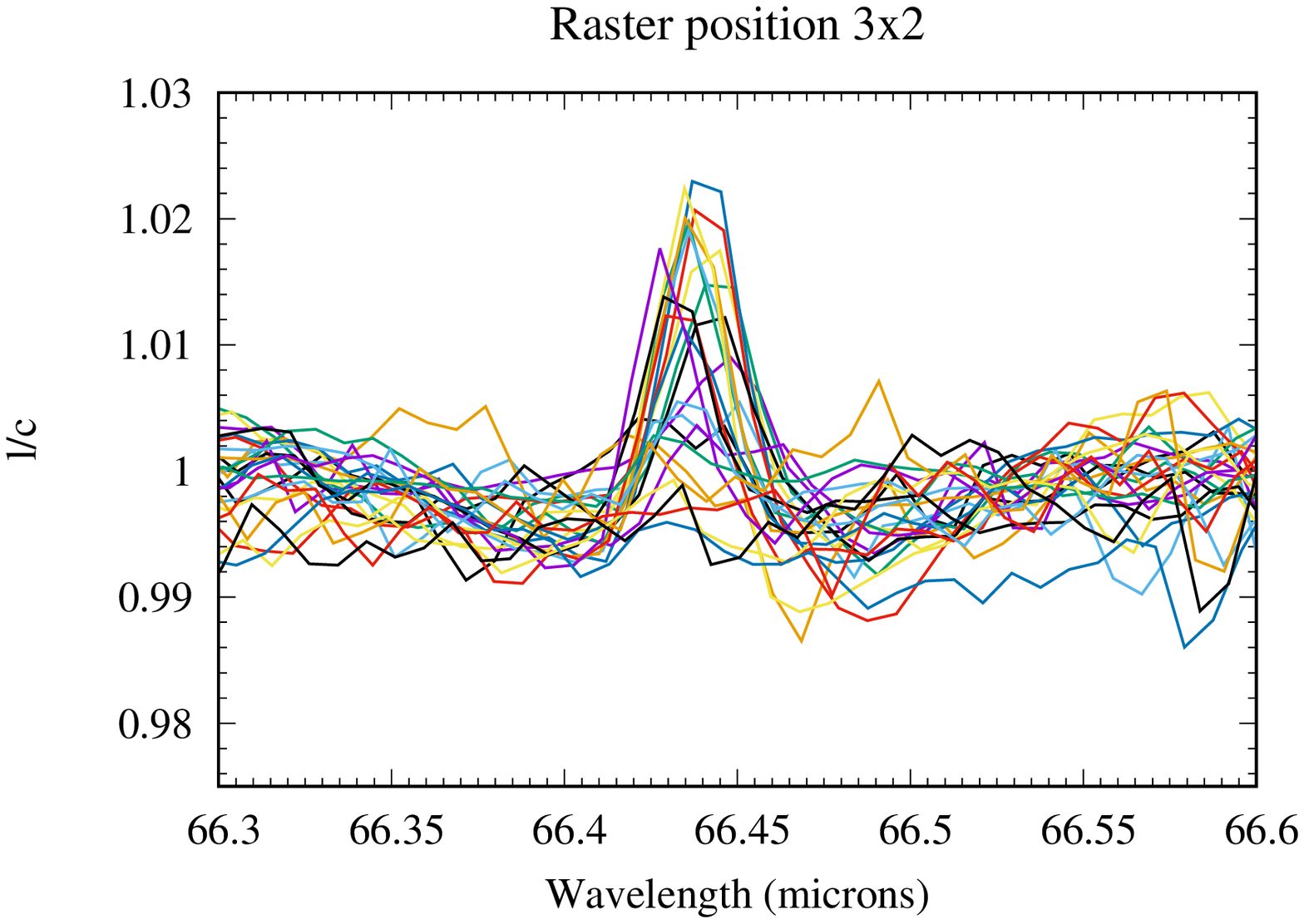}
   \includegraphics[width=7.5cm,keepaspectratio]{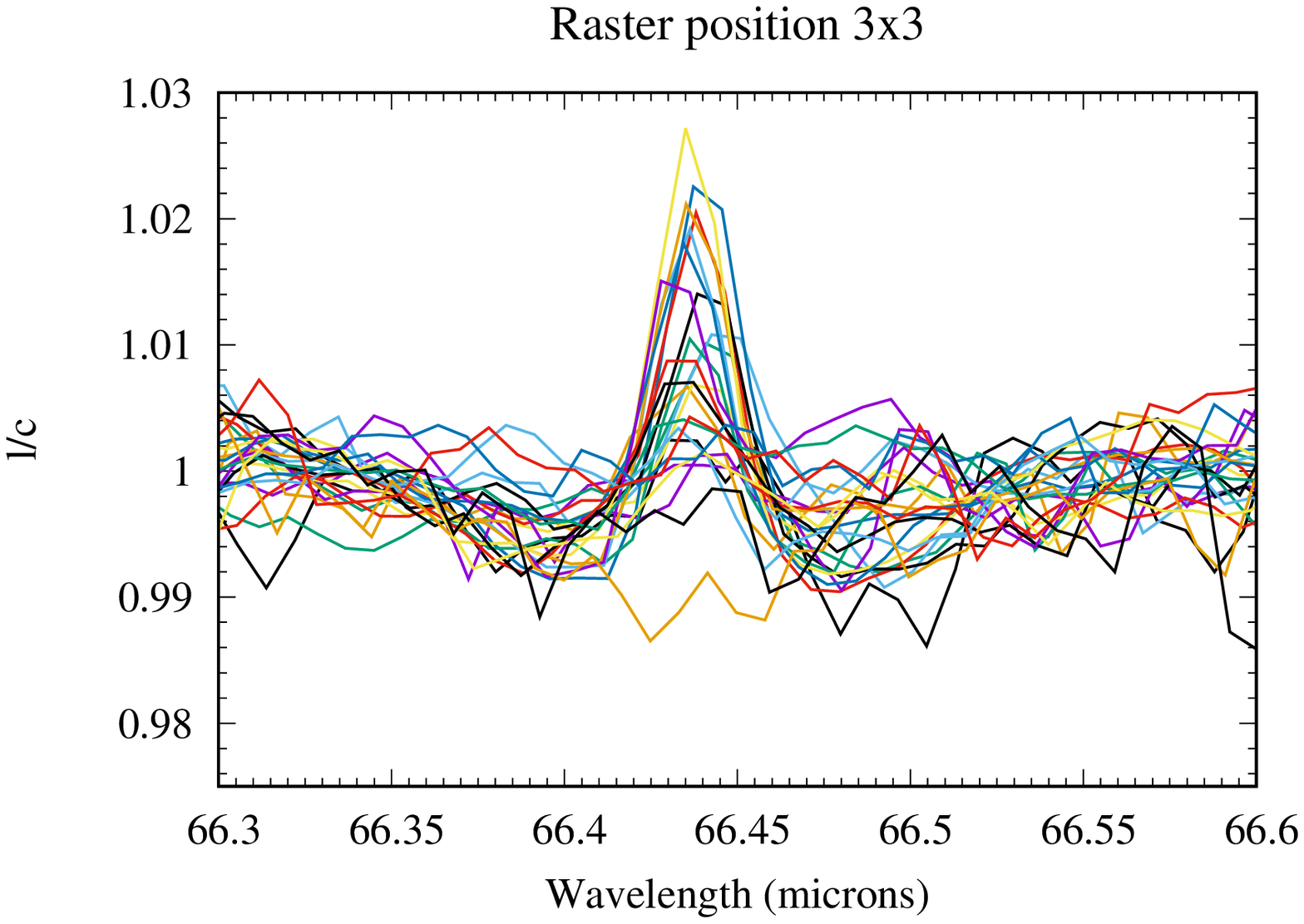}
\end{center}
\caption{225 raw spectra obtained with the PACS 5$\times$5 detector array on all of the 3$\times$3 raster map positions on Saturn at 66.44\,$\mu$m. The 25 spectra recorded by the detector array are plotted for each of the 9 raster map positions. The H$_2$O line is detected in all spectra corresponding to pointings on the planetary disk or close to the planetary limb. Spectra without an H$_2$O line correspond to poitings far off the planetary limb. Before fitting the lines with a Gaussian to compute their area, a polynomial baseline was removed from these spectra. The line peak S/N around the limb is $\sim$15-45. }
\label{raw_spectra} 
\end{figure}
\end{landscape}

\begin{figure}[!h]
\begin{center}
   \includegraphics[width=10cm,keepaspectratio]{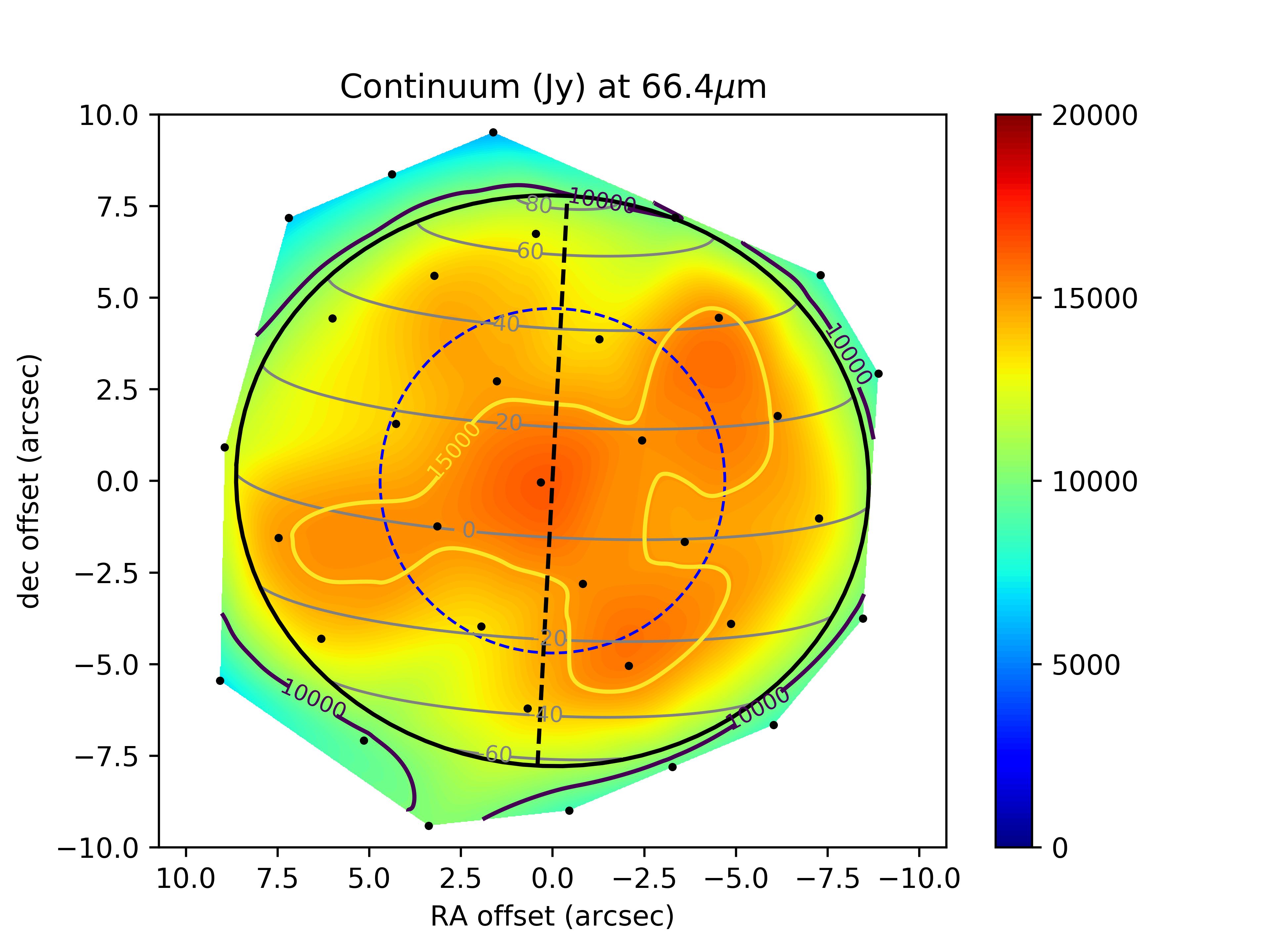}
   \includegraphics[width=10cm,keepaspectratio]{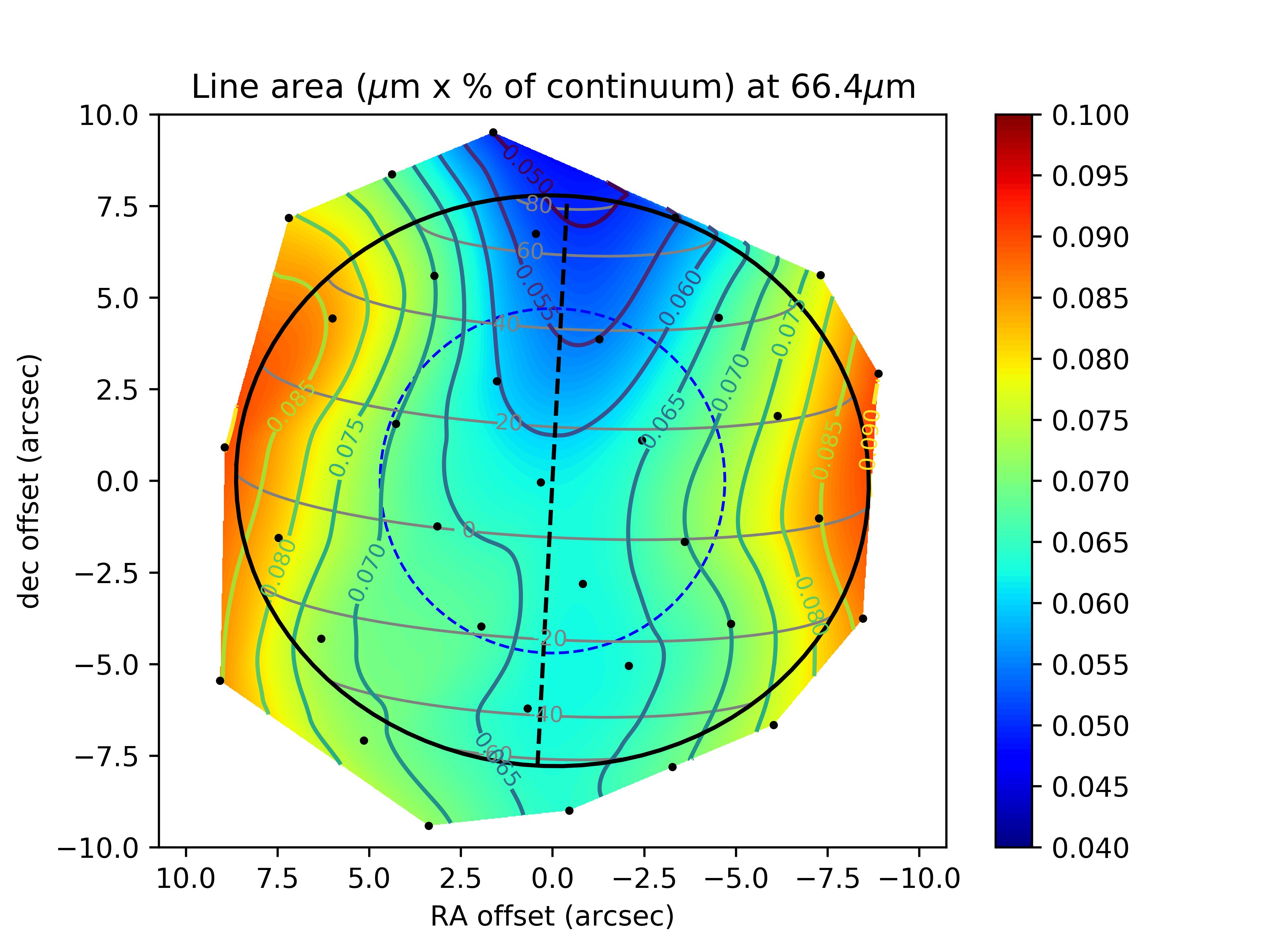}
\end{center}
\caption{Water map at 66.44\,$\mu$m observed by the PACS spectrometer on January 2, 2011. Saturn is represented by the black ellipse, and its rotation axis is also displayed with a black dashed line. Iso-latitudes are plotted with grey lines. The beam is represented by a blue dashed circle. Each black dot represents the central position of a pixel of the raster map. (Top) Image of the continuum (in Jy), after the residual pointing offset was corrected. (Bottom) The line area in microns $\times$ \% of the continuum is displayed. The S/N in this map varies from $\sim$15 to $\sim$45 (the 1-$\sigma$ rms is 0.0023 microns $\times$ \% of the continuum), and enables us to demonstrate the difference of water emission between the poles and equatorial region despite the limited spatial resolution. The map shows a lack of emission at high northern latitudes, and to a lower extent in the high southern latitudes as well.}
\label{66micron_map} 
\end{figure}

   The basic data reduction was run with HIPE 8.0 (Herschel Interactive Processing Environment, \citealt{Ott2010}) and additional processing (flat-fielding, outlier removal and rebinning) was performed with standard IDL tools, providing us with the 225 spectra presented in \fig{raw_spectra} for all raster positions of the map. After subtracting the coordinates of Saturn given by the JPL/Horizon ephemeris, we find from the continuum distribution that the map has a residual pointing offset of 2.5\arcsec~in RA and 0.3\arcsec~in dec. The position of the line peak presents some scatter, which results from a combination of two effects: (i) a Doppler shift caused by the rapid rotation of the planet,  and (ii) wavelength shifts induced by the non-uniform illumination of the instrument slit on some of the raster positions \citep{Poglitsch2010}. A baseline, caused by standing waves excited by the strong continuum emission of the planet in the instrument, was then removed from the raw spectra by fitting a polynomial. This step introduces on average a 10-15\%~uncertainty on the line amplitude. The H$_2$O line is detected in all positions within the disk and on positions within $\sim$one beam from the planetary limb. For the reasons detailed in \citet{Cavalie2013}, reasonable uncertainties cannot be achieved on an absolute flux calibration and we have to express the spectra in terms of line-to-continuum ratio ($l/c$) to cancel out the absolute response. In addition, the line width is purely instrumental (and Gaussian) and can vary from one position to another because of varying beam filling factors on the detector array. Therefore, the map must be analyzed in terms of line area. We fit the lines with a Gaussian and compute their area from the Gaussian fit parameters. According to the noise on the line peak intensity and line width, we estimate that the line-line-area maparea map can be safely analyzed for positions within the planetary disk or in the vicinity of the limb (see \fig{66micron_map}). We find a 1-$\sigma$ rms of 0.0023 (in units of microns $\times$ \% of the continuum) when adding quadratically the baseline removal uncertainty and the spectral noise. We use this value in our $\chi^2/N$ calculations ($N$ is the number of pixels in the map). The average S/N (signal-to-noise ratio) around the limb, i.e., where the lines have the biggest contrast, is $\sim$15-45 on the line area, with maximum value at the eastern and western limbs and a minimum at the northern limb. 
At a given latitude, the increased emission at the limb (compared to the disk center) is a geometrical effect, caused by limb-brightening in the line and limb-darkening in the continuum. There is enhanced emission at low latitudes with respect to mid-to-high latitudes, and a local minimum around the North Pole. The resulting continuum and line-area maps are presented in \fig{66micron_map}.




   \subsection{HIFI observations}
\begin{figure}[!h]
\begin{center}
   \includegraphics[width=8cm,keepaspectratio]{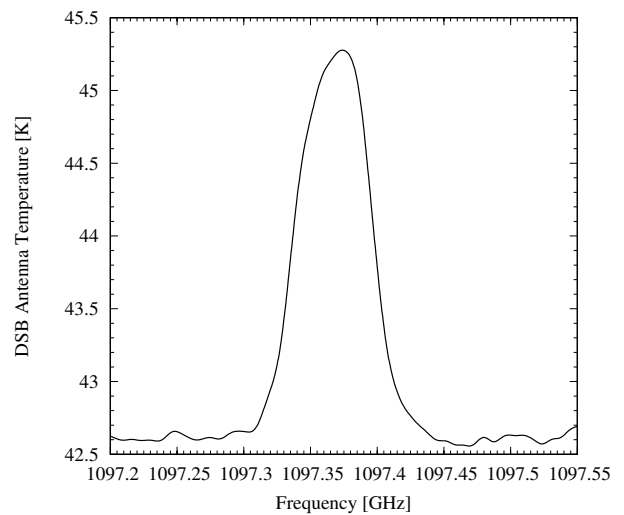}
\end{center}
\caption{Water line at 1097.365\,GHz observed by the HIFI spectrometer on December 31, 2010. The spectrum is expressed in terms of kelvins on the DSB scale. The asymmetry seen in the line shape is caused by a pointing offset of 1.5\arcsec~in the planet western limb direction. The North-South pointing offset is of the order of a few arcsec, but cannot be constrained further because of calibration uncertainties. The total uncertainty on the line-to-continuum ratio is 3\%.}
\label{HIFI_line} 
\end{figure}

   In addition, and in an attempt to check the PACS results at the planetary scale, we use the disk-averaged observation of Saturn conducted with the HIFI instrument \citep{deGraauw2010} at 1097\,GHz on December 31, 2010, with a dual beam-switch mode \citep{Roelfsema2012}, which enables pointing to the deep sky for calibration by moving the secondary mirror instead of the primary. We have chosen this strong H$_2$O line because it is not affected by the Enceladus torus absorption which is seen only in lines like the 557\,GHz and 1113\,GHz lines \citep{Hartogh2011} that have a low lower state energy relative to the ground state (23.7944 and 0.0\,cm${-1}$, respectively); the lower state energy of the 1097\,GHz line is 136.7617\,cm$^{-1}$. Given the \textit{Herschel} HPBW is 19.3\arcsec~(i.e., bigger than Saturn) and the spectral resolution of 1.1\,MHz of the Wide Band Spectrometer (WBS), the line appears smeared because of the fast rotation of Saturn (9.9\kms~at the equator). We processed the data with the standard HIPE 8.2.0 pipeline \citep{Ott2010} up to level 2 for the H and V polarizations. We then remove standing waves by using a \citet{Lomb1976} algorithm before averaging the two polarizations and smoothed the spectral resolution to 10\,MHz. This last operation barely changes the line shape as the width of the line is already $\sim$35\,MHz because of the aforementioned rotational smearing. The resulting line is displayed in \fig{HIFI_line}, and we note a small asymmetry and the fact that the line-center frequency is not aligned with the line rest frequency. This asymmetry is caused by a pointing offset of 1.5\arcsec~in the planetary western limb direction that we account for in what follows. We performed no absolute calibration and thus cannot constrain the North-South pointing offset, which must be of the order of a few arc seconds according to the observatory pointing uncertainty at this frequency \citep{Roelfsema2012}. We account for the double sideband (DSB) response of the instrument by assuming a normalized sideband ratio $G_{USB}(H+V)=0.469\pm0.016$ \citep{Higgins2014} and a ratio between the continuum in the upper and lower sidebands of 1.037 (according to our continuum model), and produce a single sideband (SSB) spectrum that we use for our analyses in terms of line-to-continuum ratio ($l/c$) in this paper. The total uncertainty on the line-to-continuum ratio, when accounting for the uncertainties caused by the normalized sideband ratio, the baseline removal process, and the spectral noise, is 3\%, i.e., 0.08\,K. All our $\chi^2/N$ calculations are based on this rms value, which does not include absolute calibration uncertainties that are estimated to be 5\%.

\section{Modeling \label{Models}}
In this section, we present the thermal and radiative-transfer models that have been applied in our data analysis. 

   \subsection{Thermal model}
\begin{figure}[!h]
\begin{center}
   \includegraphics[width=10cm,keepaspectratio]{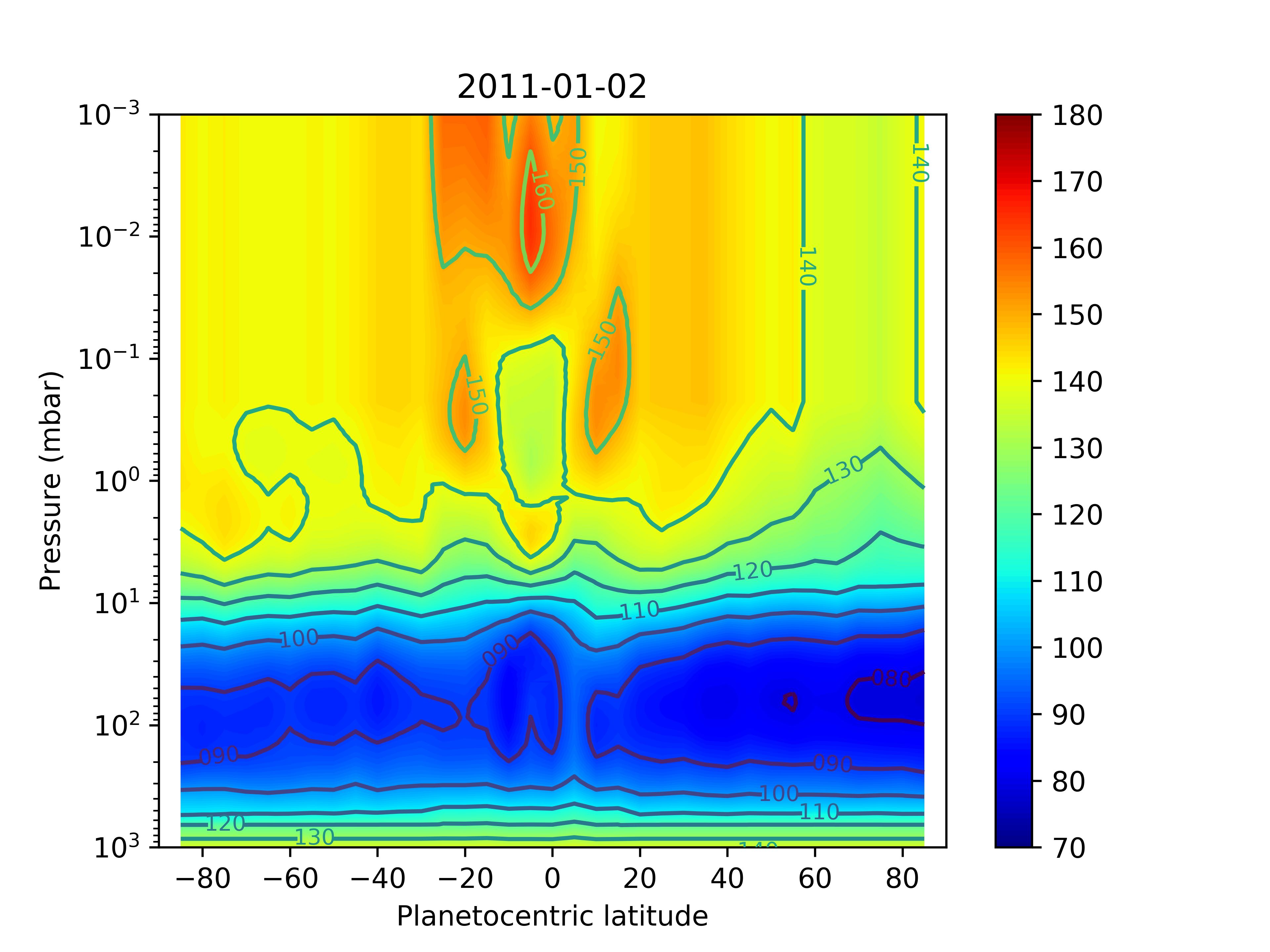}
   \includegraphics[width=8.5cm,keepaspectratio]{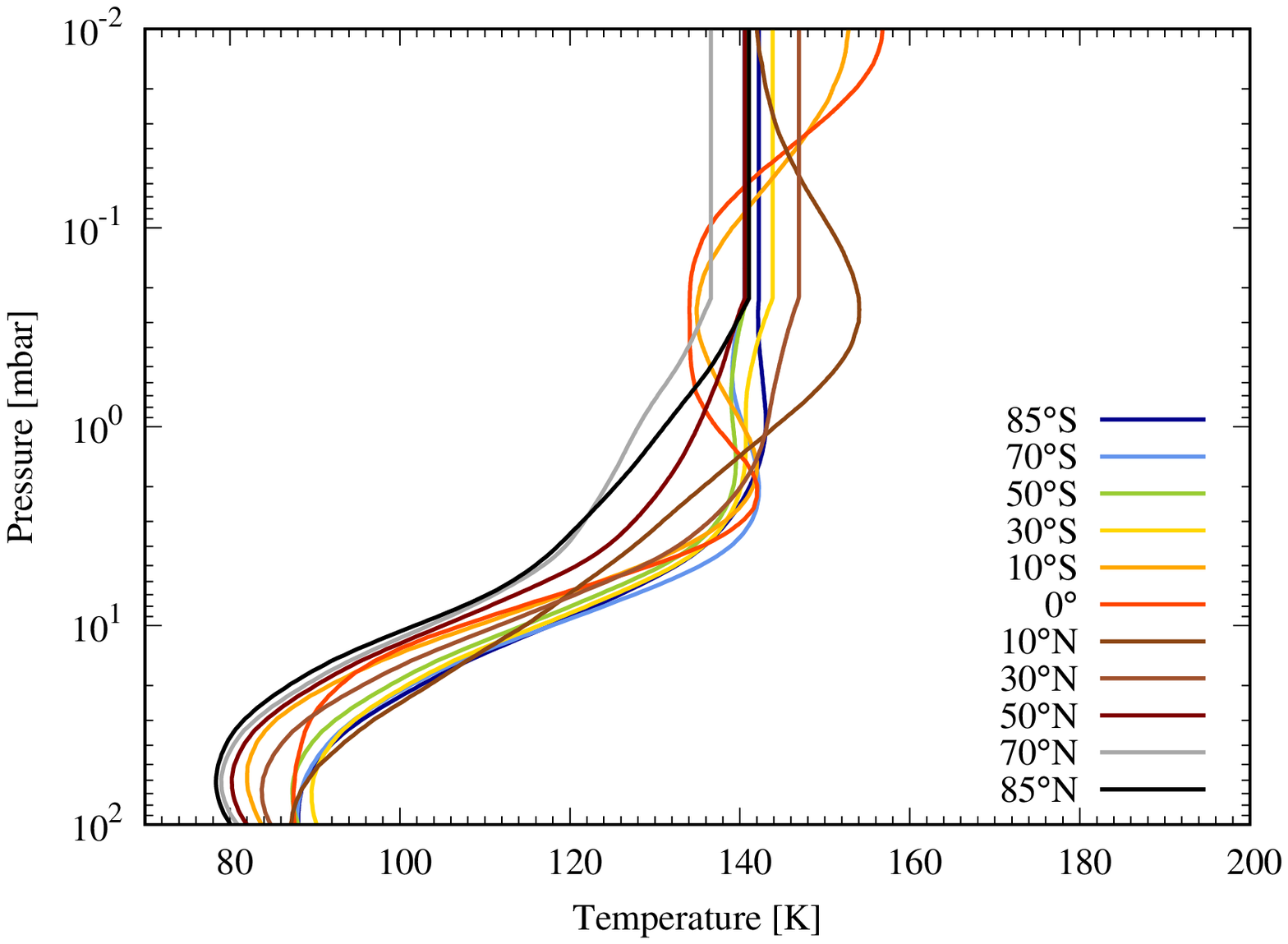}
\end{center}
\caption{(Top) Nominal thermal field used in this paper. Zonally-averaged temperatures are given as a function of pressure and planetocentric latitude. The data combine retrievals from nadir and limb Cassini/CIRS data (see text for references). (Bottom) Temperature vertical profiles extracted from the field for a set of latitudes. }
\label{Thermal_field_prestorm} 
\end{figure}

   The pressure-latitude thermal field we use in our radiative-transfer modeling is extracted from the spline interpolation made over time to fill in missing parts of Cassini's Composite Infrared Spectrometer (CIRS) time series spanning the full mission \citep{Fletcher2017}. We can thus extract the zonally-averaged stratospheric temperatures pertinent to the dates of the Herschel observations. At the time of our observations, the CIRS dataset features significant temperature anomalies associated with the planetary storm that erupted at the end of 2010 \citep{Sanchez-Lavega2011,Fischer2011,Fletcher2011}. This storm significantly changed the northern stratospheric temperatures in a 20\degre~latitude band centered on 40\degre N planetocentric. Two hot stratospheric vortices formed well above the tropospheric storm and were nicknamed ``beacons'' because of their bright appearance in the IR and the rapid rotation of Saturn. They eventually merged in April-May 2011 to form a huge vortex in which a temperature increase of up to 80\,K was measured at the 2-mbar level. But, because our observations occur quite early in the time evolution of the storm and beacons, we first choose to produce a thermal field representative of storm-free conditions by removing any coverage between 20N-60N and 2011-2012 in the spline smoothing, and will include the beacons later in the paper (in Section~\ref{Storm}). We take the thermal field produced for the date closest to our observations, i.e., December 28, 2010. The stratospheric temperatures in the 0.1-10\,mbar range, i.e., roughly the range in which the HIFI and PACS lines are sensitive (see \fig{Contribution} right), are quite symmetric in latitude with respect to the equator up to $\sim$40\degre, with a gradient from 40S to the South Pole $<$5\,K and a $-$10\,K gradient from 40N to the North Pole, as already hinted for observations at similar $L_S$ by \citet{Sinclair2013}. The CIRS observations used 600-1400\,cm$^{-1}$ in nadir geometry, therefore probing only the 0.2-5\,mbar and 70-250\,mbar ranges. Outside of these ranges, the profile goes back to an a priori profile built by averaging \citet{Guerlet2009,Guerlet2010} limb observations, which probe higher stratospheric altitudes than the study of \citet{Fletcher2017}, between latitudes of 45N and 45S. However, whether this a priori is valid for all latitudes is questionable. Therefore, we choose to extrapolate isothermally the temperature profiles for pressures lower than 0.2\,mbar. Between 20S and 20N, we use the thermal profiles of February 2010 derived by \citet{Guerlet2011} from Cassini/CIRS limb viewing geometry observations in order to benefit from the higher vertical resolution of such observations. With such data, the equatorial quasi-periodic oscillation \citep{Fouchet2008,Orton2008,Guerlet2011} is better resolved than in the nadir data, and the vertical sensitivity in the temperature retrieval is extended to 10$^{-2}$\,mbar. \fig{Thermal_field_prestorm} displays the thermal field. We note that the temperatures in the submillibar range are in agreement with temperatures derived from Voyager 2 occultation experiments at several latitudes by \citet{Lindal1985} for a similar $L_S$. 

   \subsection{H$_2$O distribution empirical models}
   In this paper, we test two types of H$_2$O spatial distributions that are representative of two different sources: IDPs and the Enceladus plumes.
   
   The IDP source is modeled with a meridionally uniform distribution of H$_2$O. The mole fraction of H$_2$O is uniform above the local condensation level, which is computed following \citet{Fray2009} at each latitude.
   
   The Enceladus source is modeled with a 2-parameter Gaussian-shaped latitudinal distribution of H$_2$O, centered on the equator, with a peak H$_2$O mole fraction at the equator $y_\mathrm{eq}$, and a half-width at half-maximum (HWHM) $\sigma$, such that:
   \begin{equation}
   y_{\mathrm{H}_{2}\mathrm{O}}(\phi)=y_\mathrm{eq}\times \exp\left(-\frac{\phi^2}{2\sigma^2}\right) \label{H2O_gaussian}
   \end{equation}
   where $y_\mathrm{eq}$ is the H$_2$O mole fraction above the condensation level at the equator, and $\phi$ is the planetocentric latitude. Here also, latitudinally-dependent condensation is computed from our thermal field. The recently observed ring source \citep{Waite2018,Hsu2018,Mitchell2018} shares common properties with the Enceladus source: it is centered on the equator and has, to the first order, a Gaussian shape. 
   
   The ring atmosphere (mostly O$_2$, not H$_2$O), another potential source for Saturn's stratospheric H$_2$O, has lower densities than the Enceladus neutral torus \citep{Johnson2006,Cassidy2010}. Neutral-neutral collisions are therefore inefficient, and even ion-neutral charge exchange collisions are slow. Unlike elsewhere in the magnetosphere, the co-rotating plasma flow speed is similar to the orbital speed. The ring atmosphere is therefore lost at higher latitudes like ring rain: the atmosphere is lost via photoionization followed by precipitation along field lines \citep{Moore2015}. We note that the neutral component of their H$_2$O distribution is actually tied to Enceladus. This is why we do not consider the ring atmosphere as a significant enough source to explain our observations, and will not include it in our models.

   \subsection{Radiative-transfer model \label{RT_model}}
   We model the observations with a line-by-line non-scattering model that solves the radiative-transfer equation in 1D on several thousand lines-of-sight, thus highly oversampling our observational spatial resolution. We use the spectral line parameters from the JPL Molecular Spectroscopy Database \citep{Pickett1998}. The spectra of collision-induced absorption caused by H$_2$-H$_2$, H$_2$-He, and H$_2$-CH$_4$ pairs are included following \citet{Borysow1985,Borysow1988} and \citet{Borysow1986}. More recent updates to the H$_2$-H$_2$ absorption from ab initio quantum theory exist \citep{Orton2007,Fletcher2018a}, but they do not make any significant difference in this spectral region. We adopt mole fractions of 11.8\% and 0.47\% for He and CH$_4$ respectively, according to \citet{Conrath2000} and \citet{Fletcher2009a}. We account for the absorption caused by the H$_2$O line, as well as surrounding NH$_3$ and PH$_3$ broad lines. We take the NH$_3$ and PH$_3$ distributions from \citet{Davis1996} and \citet{Fletcher2009b}, respectively. We compute the relevant H$_2$/He broadening parameters from \citet{Levy1993,Levy1994} for PH$_3$, from \citet{Fletcher2007} for NH$_3$, and from \citet{Brown1996} and \citet{Dick2009} for H$_2$O, for the H$_2$/He mixture described above and the appropriate temperature range. 
The 1097\,GHz line has an opacity of $\sim$0.3 over the HIFI beam in disk-centered conditions, and the 66.44\,$\mu$m line has an opacity that ranges from 0.1 to 0.4 over the PACS beam for disk-centered and limb conditions, for the best fit model H$_2$O profiles given in \fig{abundance_gradient7}.

We note that, contrary to the 66.4\,$\mu$m line seen by PACS, the 1097\,GHz line seen by HIFI lies on the far wings of a PH$_3$ line. A change in the PH$_3$ abundance therefore influences the line-to-continuum ratio of the H$_2$O line and subsequently on the derived H$_2$O abundance. Increasing the PH$_3$ abundance by a factor of 2 decreases the H$_2$O abundance required to fit the line by $\sim$15\%.    
   

   This model significantly improves on the model already described in \citet{Cavalie2008a,Cavalie2013} regarding the handling of the geometry. The observation plane is sampled with an irregular grid so that the limb and rings, i.e., the regions with the highest variations in terms of continuum and line emission, are sufficiently sampled without hampering the computational efficiency. We account for the 3D ellipsoidal geometry of Saturn and rings at the time of the observations, using the parameters listed in \tab{Obs_list}. This means we compute the local latitudes, longitudes, and altitudes, at each sampled location on each line-of-sight, and are thus able to use thermal and abundance fields that are fully 3D, before applying the beam convolution. 
   
   In January 2011, the rings had an inclination of 12.4\degre, as seen from \textit{Herschel}. We thus account for the contribution of the rings to the continuum emission. The A-B-C ring system brightness temperatures are computed in the wavelength range of our observations by interpolating the observational results obtained in the infrared with Cassini/CIRS \citep{Flandes2010} and at 2\,cm with the Very Large Array \citep{vanderTak1999}, accounting for the brightness temperature roll-off seen around 200\,$\mu$m by \citet{Spilker2003,Spilker2005}. We use the solar elevation as seen from the rings given in \tab{Obs_list}, and the optical depths of the different rings from \citet{Altobelli2014} and \citet{Guerlet2014}. Results are in general agreement with \citet{Dunn2005}, when considering averages for the whole ring system. In practice, we identify geometrically the lines-of-sight that cross the rings and we account for the ring absorption and thermal emission, once all the lines-of-sight on the planet and limb are computed. Beyond the planetary disk, we find the positions of the rings and add their thermal emission. We then create a regular grid of lines-of-sight from which we compute the beam-averaged spectra, in such a way that the beam is highly oversampled by the grid. We produce maps of line integrated emission (for PACS) and disk-averaged spectral line (for HIFI) that can directly be compared with the observation. We present the simulations as well as the residuals resulting from the subtraction of the model from the observations\footnote{In the case of the HIFI spectrum, which was not absolutely calibrated, we have rescaled the spectrum to match our model continuum. The analysis is then done in terms of line-to-continuum ratio.}.

\section{Results \label{Results}}
   \subsection{Meridionally uniform distribution}
\begin{figure}[!h]
\begin{center}
   \includegraphics[width=10cm,keepaspectratio]{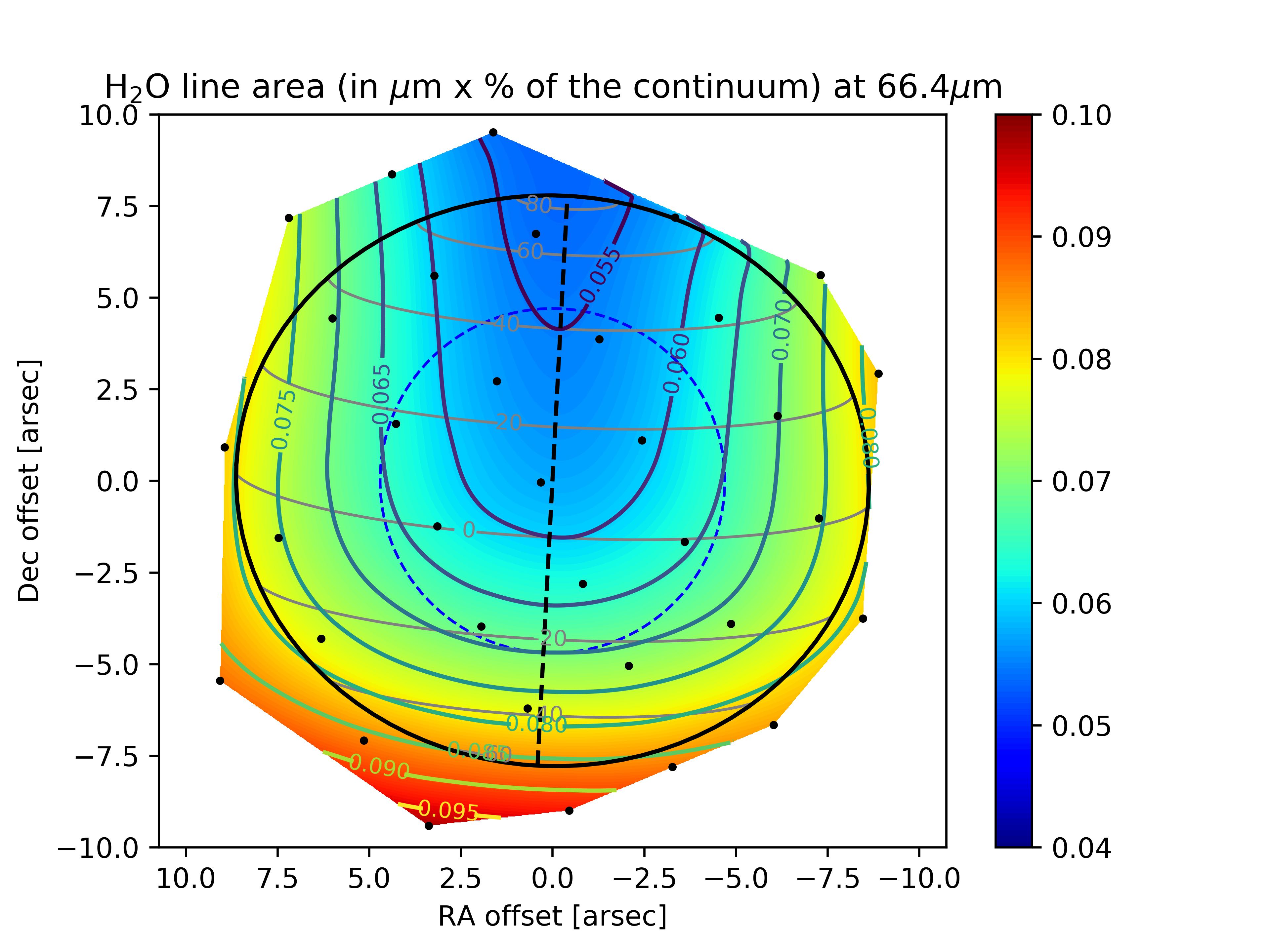}
   \includegraphics[width=10cm,keepaspectratio]{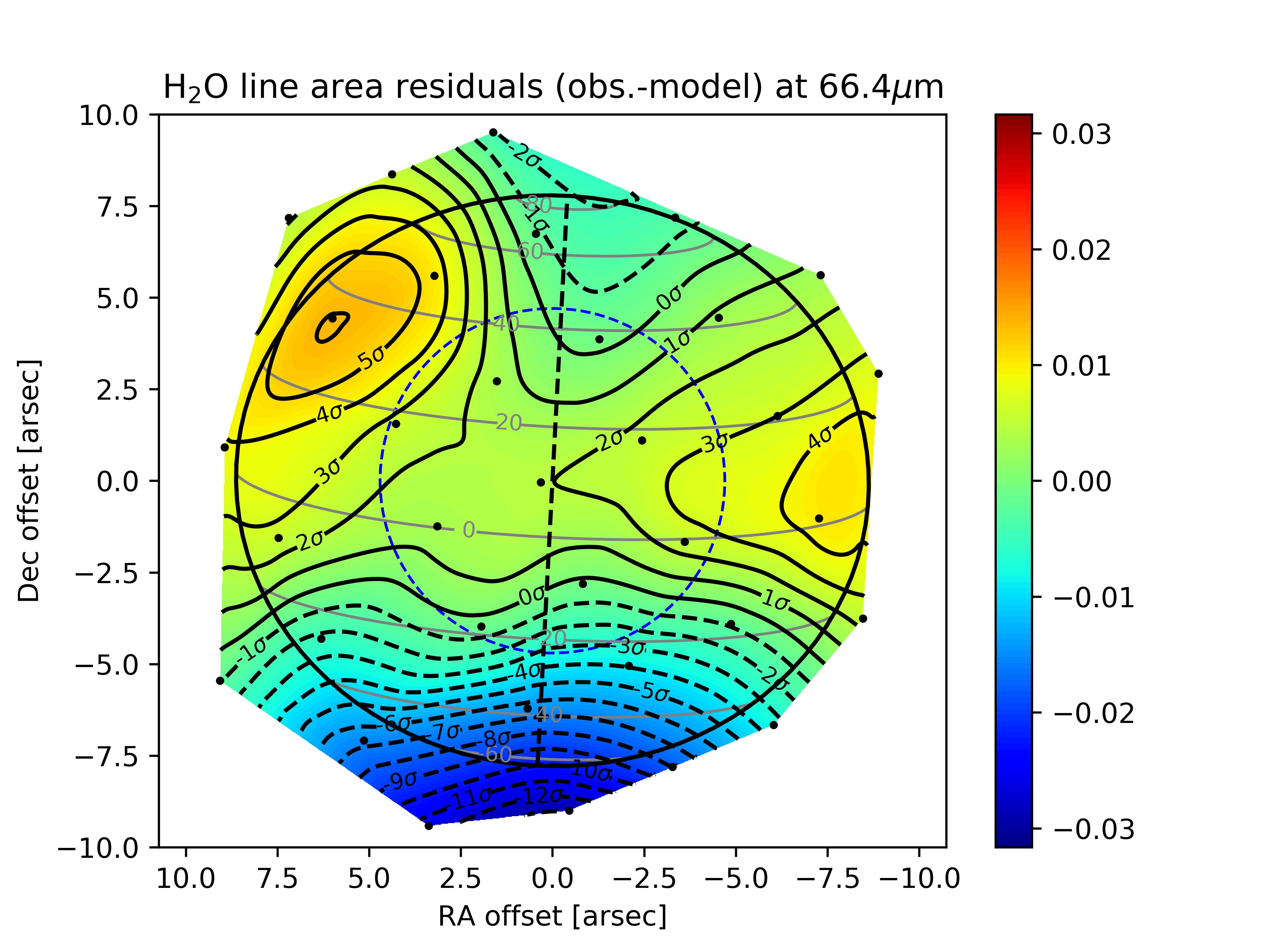}
\end{center}
\caption{(Top) Line-area map at 66.44\,$\mu$m modeled with a meridionally uniform distribution of H$_2$O. The line area is expressed in microns $\times$ \% of the continuum and can thus be directly compared to the observed map of \fig{66micron_map}. (Bottom) Map of the residuals between observations and model (in the same unit as the line-area map). Positive values indicate an excess of emission in the data compared to the model, while negative values indicate a lack of emission compared to the model. The contours are labelled with respect to the 1-$\sigma$ noise level measured in \fig{66micron_map}. Above the condensation level, the H$_2$O mole fraction is set to 6\dix{-10}. Neither the low latitudes nor the Southern latitudes are within 3-$\sigma$.}
\label{PACS_Uniform} 
\end{figure}

   An initial approach was to try to fit the PACS map with a uniform distribution of H$_2$O which would be representative of an IDP source. As shown in \fig{PACS_Uniform}, the H$_2$O line-area map is a direct translation of the thermal field and we find a strong gradient between the North and South poles. The minimum $\chi^2/N$ is obtained for an H$_2$O mole fraction of 6\dix{-10} above the condensation level (roughly a few mbar, varying with latitude), but the $\chi^2/N$ is as high as 18.9 (i.e., more than 4-$\sigma$ away from the data). The observed decrease of the line area around the southern limb is not reproduced (see \fig{PACS_Uniform}), as a results of the warmer stratospheric temperatures in the South compared to the North (see \fig{Thermal_field_prestorm}). On the contrary, the strongest emission comes from the south when we use a meridionally uniform distribution of H$_2$O. In addition, there is not enough emission at low latitudes. Such a solution is thus not satisfactory and can be discarded at this stage. 
   
   We note that an acceptable fit to the disk-averaged HIFI data can be obtained for a uniform mole fraction of $\sim$3.5\dix{-9}, which is consistent with \citet{Fletcher2012} but incompatible with the PACS data. When we simulate the H$_2$O line at 556.936\,GHz with this distribution, we obtain a line-to-continuum ratio that is consistent with the previous observation of \citet{Bergin2000}. This shows that the disk-averaged H$_2$O influx responsible for our observations of $\sim$6\dix{5}\Fluxunit~\citep{Moses2017}, which translates into a disk-integrated mass flux of $\sim$8\,kg/s, is surpassed by orders of magnitude by the extraordinarily high flux recently measured by INMS of $\sim$10$^4$\,kg/s coming from the inner ring system. This proves that the ring source observed by Cassini in 2017 cannot be the cause of the H$_2$O observed by \textit{Herschel} in 2010-2011. The ring source must therefore be more recent than our observations, or at least this source had not been active for long enough (i.e., $\sim$ 10-15\,yr) to diffuse down to observable levels ($\sim$0.1-1\,mbar). According to \citet{Waite2018}, such a large flux is probably linked to the appearance of clumps of material in the D68 ringlet in 2015 \citep{Hedman2016}, and is unsustainable for more than 1\,Myr without depleting the rings. In conclusion, we will thus not consider the ring source in our subsequent modeling of the \textit{Herschel} observations.

   \subsection{Meridionally variable distributions \label{gaussian_distribution}}
   Despite the low spatial resolution of the PACS observations, the contrast observed between the eastern/western limbs and the northern/southern ones (\fig{66micron_map} bottom), combined with the knowledge of the altitude/latitude thermal field (\fig{Thermal_field_prestorm}), provides us with a good tool for deriving the meridional distribution of H$_2$O in Saturn's stratosphere. 
   
\begin{figure}[!h]
\begin{center}
   \includegraphics[width=10cm,keepaspectratio]{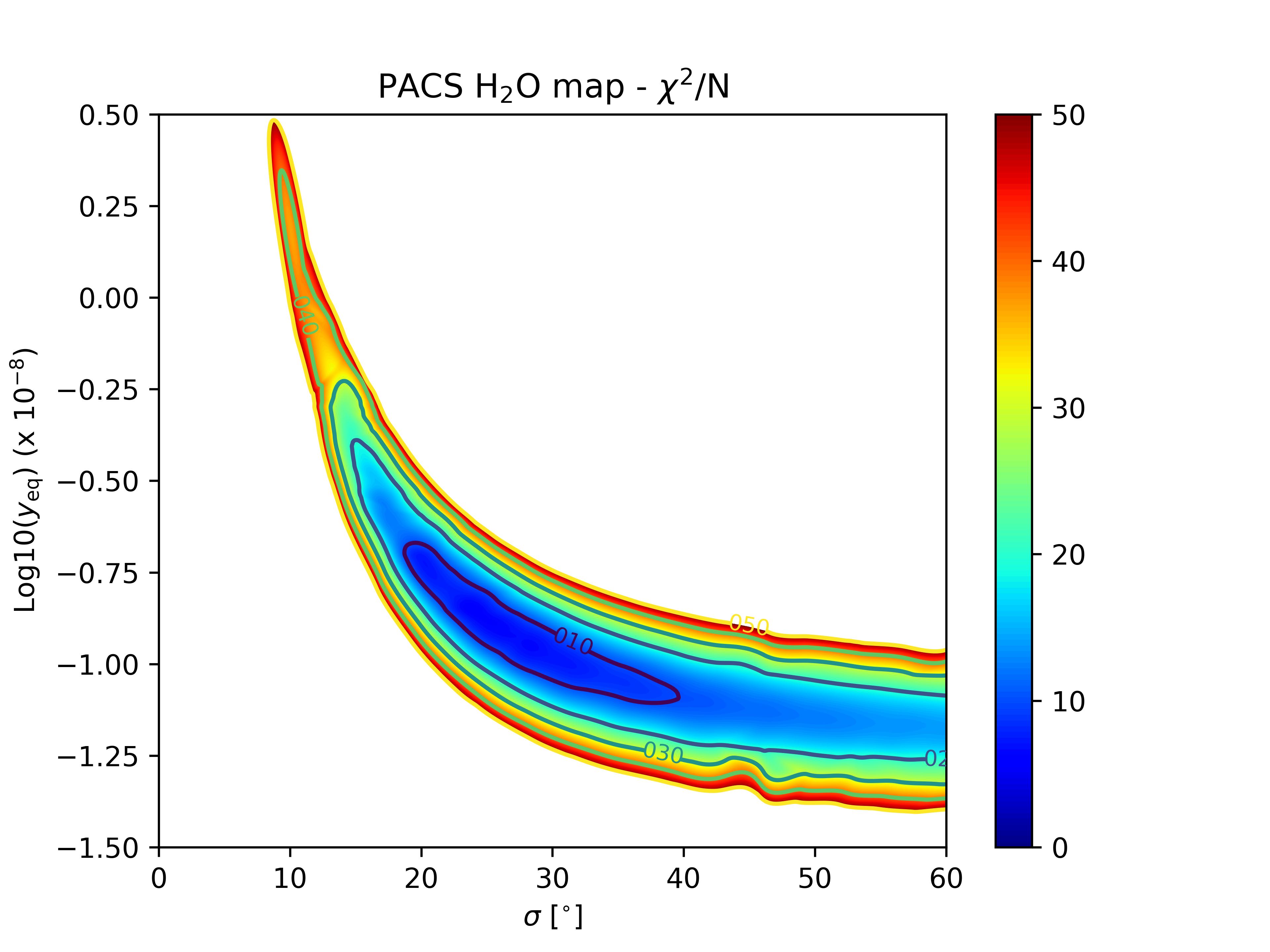}
\end{center}
\caption{$\chi^2/N$ as a function of the H$_2$O meridional gaussian distribution parameters $y_\mathrm{eq}$ and $\sigma$ (gaussian HWHM in degrees). Acceptable solutions are found for $\sigma$ ranging from $\sim$20\degre~to $\sim$40\degre, with a minimum for $\sigma$$=$25\degre~and $y_\mathrm{eq}$$=$1.4\,ppb. Corresponding line area and residual maps are shown in \fig{PACS_gaussian}.
}
\label{chi2_PACS_gaussian} 
\end{figure}

\begin{figure}[!h]
\begin{center}
   \includegraphics[width=10cm,keepaspectratio]{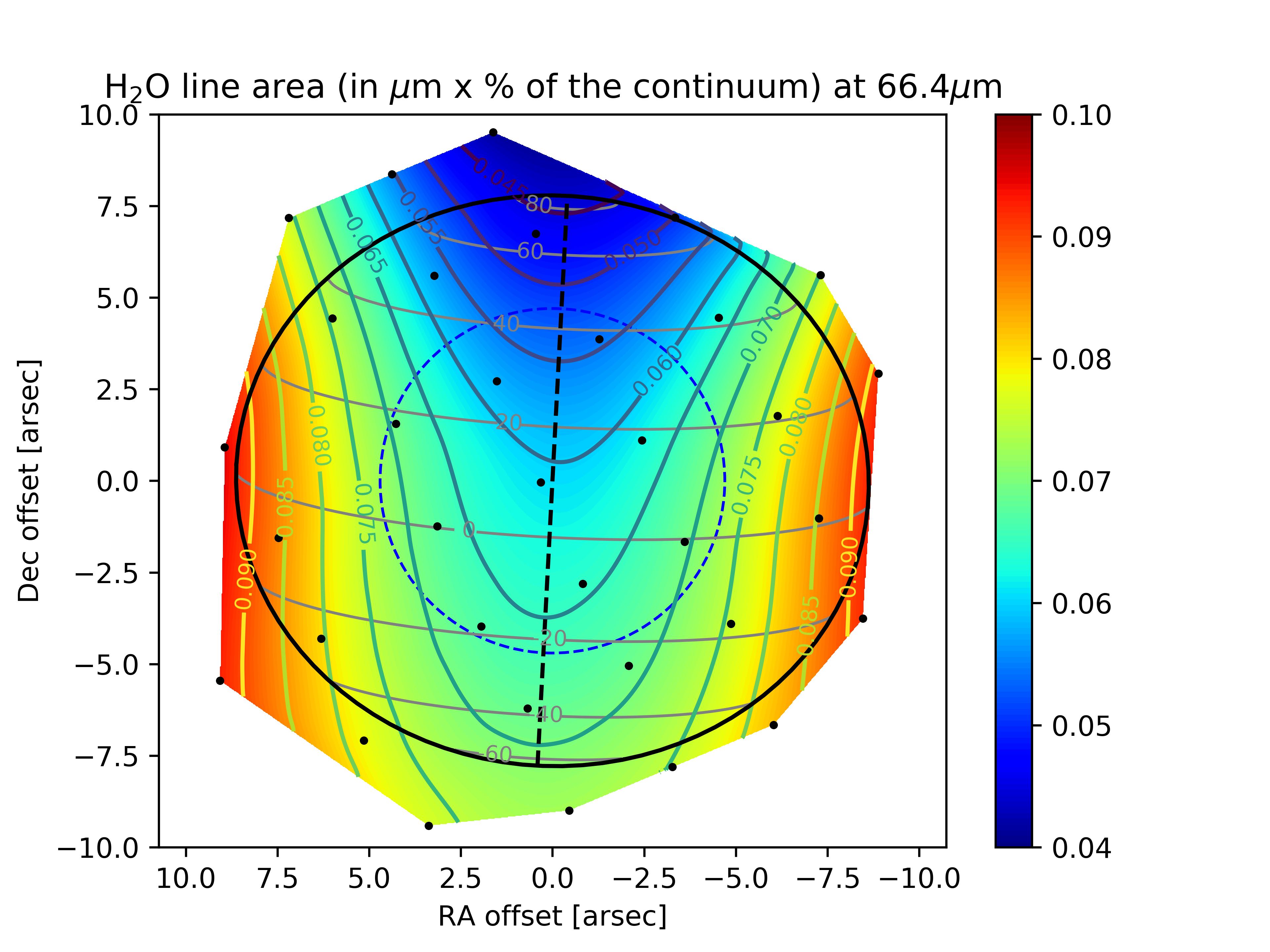}
   \includegraphics[width=10cm,keepaspectratio]{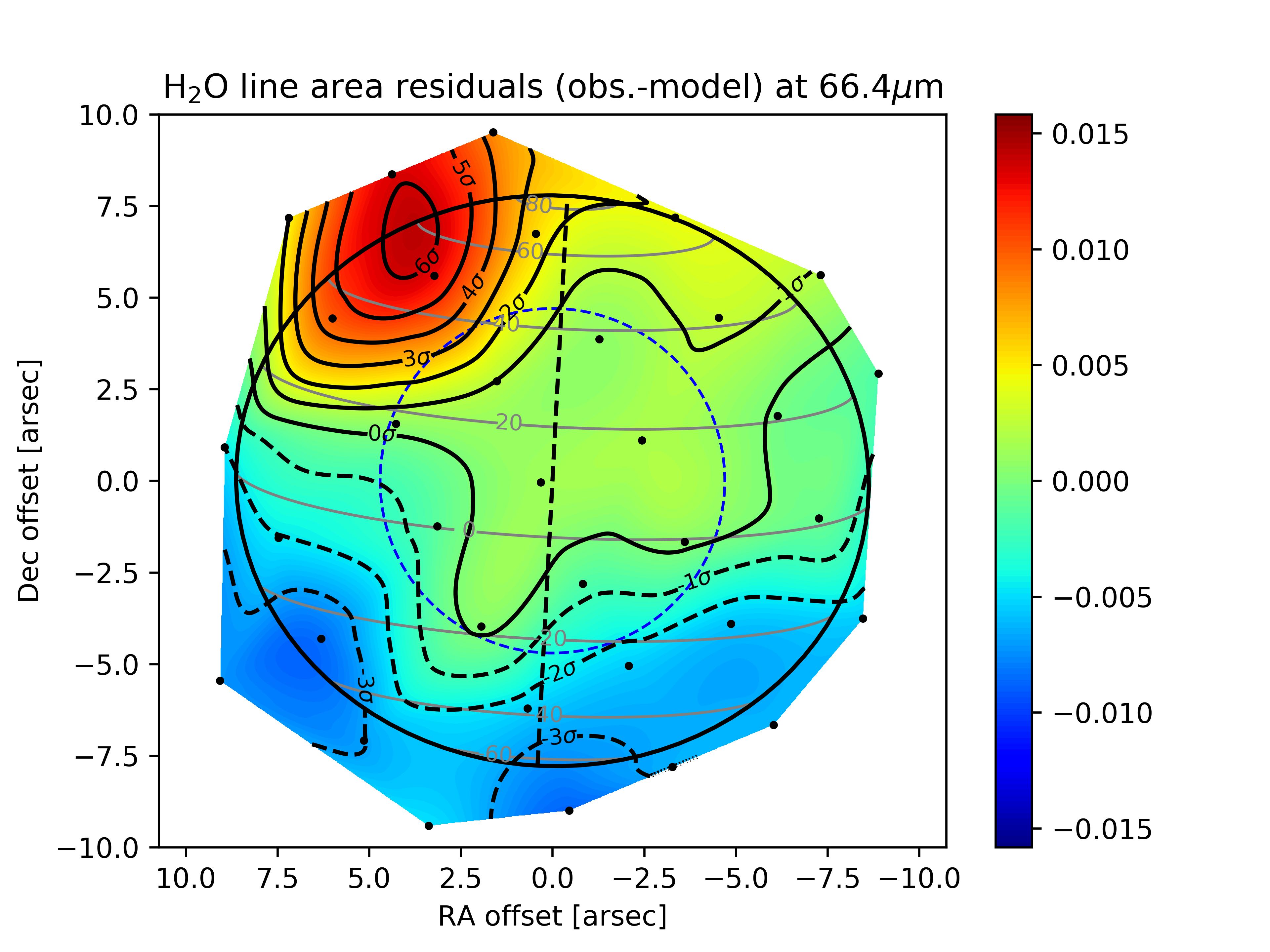}
\end{center}
\caption{Same as \fig{PACS_Uniform} for a gaussian distribution of H$_2$O around the equator with $y_\mathrm{eq}$$=$1.4\,ppb and $\sigma$$=$25\degre. The overall fit is acceptable, even if the fit is not good at the northwest limb. A better fit in this region can be obtained by increasing $y_\mathrm{eq}$, but the model would then depart even more from the data in the southern hemisphere. It is already the case in this model, which is just the best balance between the northwest and southern emission.
}
\label{PACS_gaussian} 
\end{figure}

   In what follows, we test several cases of H$_2$O meridional distributions as parametrized in eq. \ref{H2O_gaussian}. We first explore the $y_\mathrm{eq}$--$\sigma$ parameter space. 
   \fig{chi2_PACS_gaussian} shows the $\chi^2/N$ obtained as a function of these two parameters. There is a series of marginally acceptable solutions ($\chi^2/N$$<$9). The best fit is obtained for $\sigma$$=$25\degre~and $y_\mathrm{eq}$$=$1.4\,ppb, with $\chi^2/N$$=$6.3, and the corresponding line area and residual maps are displayed in \fig{PACS_gaussian}. We cannot obtain better overall fits with this model, because of a remaining excess of emission in the northwestern limb (and to a lower extent in the northeastern limb). Minimizing the residuals implies compensating partially for this excess, which results in too much emission in the southern hemisphere. 
Thus, we first try to improve the fit by reassessing our temperature field in the next section, since the temperature field at the time of the observations was influenced by the onset of Saturn's Great Storm of 2010-2011. 
      
\begin{figure*}[!h]
\begin{center}
   \includegraphics[width=9cm,keepaspectratio]{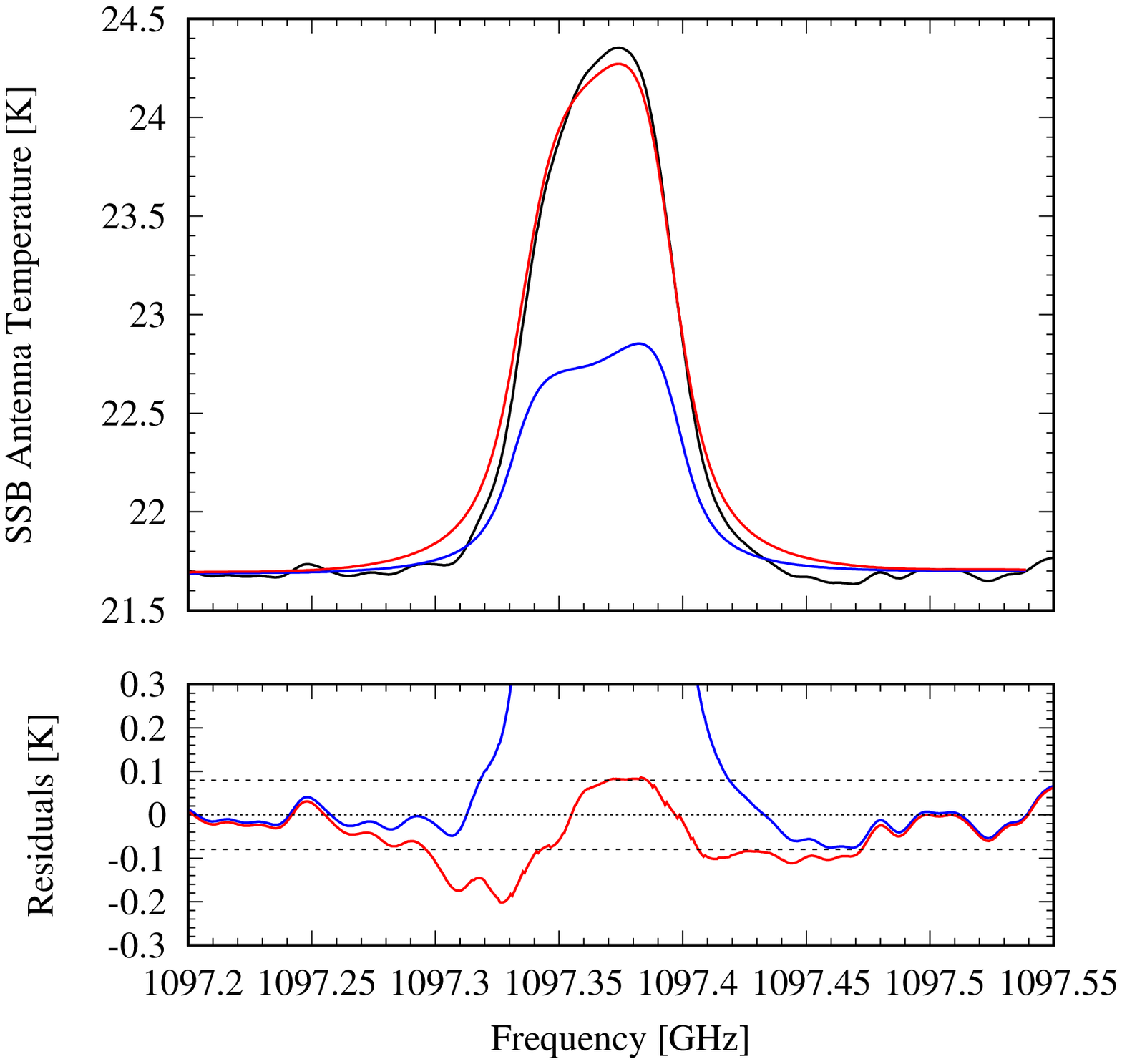}
   \includegraphics[width=9cm,keepaspectratio]{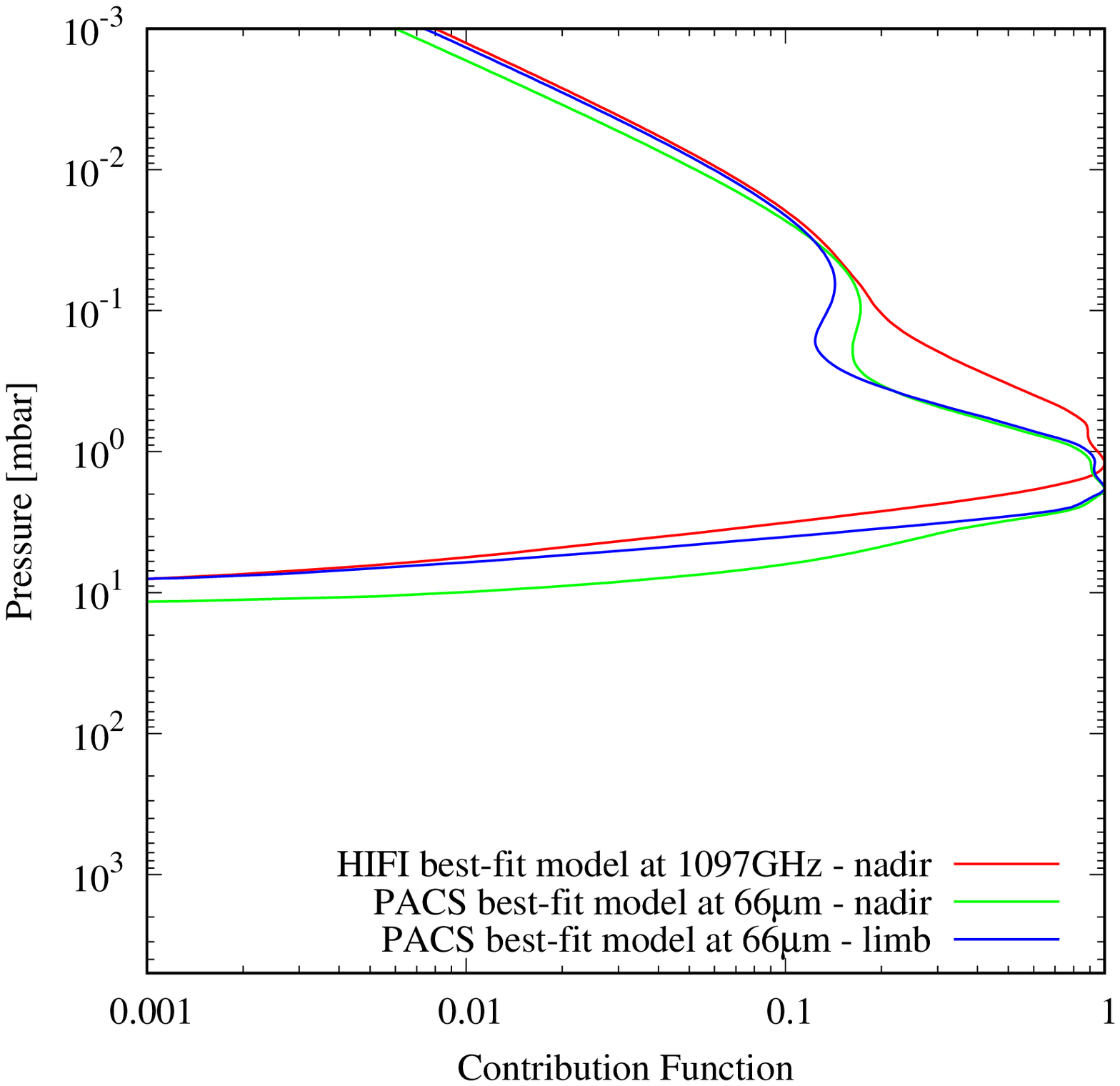}
\end{center}
\caption{(Left) Best fit to the HIFI data for a Gaussian distribution of H$_2$O around the equator with $y_\mathrm{eq}$$=$7.2\,ppb and $\sigma$$=$25\degre. The data are shown in black and the model in red, and the residuals (observation $-$ model) are plotted at the bottom with the 1-$\sigma$ level of noise in black dashed lines. The synthetic spectrum is too broad and there is a lack of emission at the center. The best-fit model to the PACS data from \fig{PACS_gaussian} is shown in blue for comparison. (Right) Normalized and beam-averaged contribution functions at the centers of the 66.44\,$\mu$m line in limb geometry (blue) and at the disk-center (green), and of the 1097\,GHz line at the disk-center (red). Both functions are computed at their respective observed spectral resolution, and the contribution of the continuum emission has been subtracted. While the PACS line essentially probes near the condensation level, the HIFI line probes slightly lower pressures. We note that we have rescaled the HIFI contribution function by multiplying it by a factor of 1.45 for an easier comparison with the PACS one.
}
\label{Contribution} 
\end{figure*}

   As in the uniform distribution case, the HIFI data requires a $\sim$5 times higher $y_\mathrm{eq}$ than the PACS data for $\sigma$$=$25\degre~to minimize $\chi^2/N$. The fit, shown in \fig{Contribution} left, is also only marginally acceptable. The line is too broad and not strong enough. This is an indication that models as simple as constant profiles above the condensation level are not optimal. The HIFI line-center contribution function peaks indeed at slightly lower pressures than the PACS line contribution function (see \fig{Contribution}). Introducing a positive gradient in the H$_2$O vertical profile above the condensation layer should thus (i) reduce the width of the HIFI line by removing some H$_2$O in the levels just above the condensation layer, (ii) increase the HIFI line amplitude by adding some H$_2$O at higher levels to which the HIFI line center is still sensitive, and (iii) help to reconcile the HIFI and PACS data as they probe different altitudes.
      
   In an attempt to improve the fit to the PACS data and to improve the compatibility with the HIFI data, we apply a two-step approach, in which we first account for the temperature changes caused by Saturn's 2010-2011 Great Storm (Section~\ref{Storm}), and then apply a parametrized gradient in the H$_2$O vertical distribution above its condensation level (Section~\ref{sec_gradient}).

   \subsection{Accounting for Saturn's Great Storm of 2010-2011 \label{Storm}}
   The \textit{Herschel} observations were conducted a few weeks after the formation of a huge storm system around 40\degre~in the northern hemisphere of the planet \citep{Sanchez-Lavega2011,Fischer2011,Fletcher2011}. Between December 2010 and April 2011, it formed a cool region surrounded by two confined warm regions in Saturn's stratosphere, referred to as ``beacons'' (B1 and B2) given their appearance in the infrared combined with Saturn's fast rotation. 
   
\begin{figure*}[!h]
\begin{center}
   \includegraphics[width=9cm,keepaspectratio]{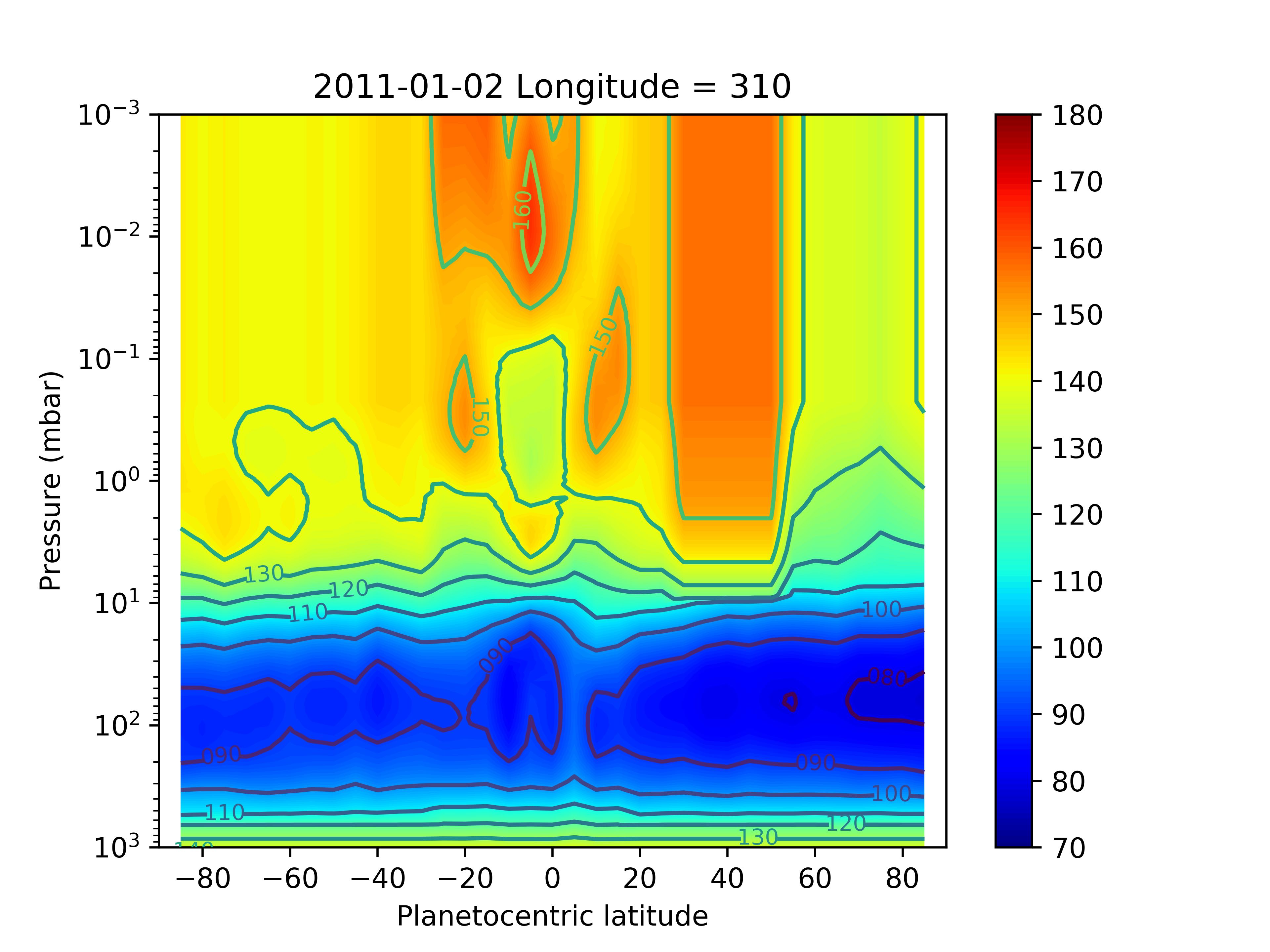}
   \includegraphics[width=9cm,keepaspectratio]{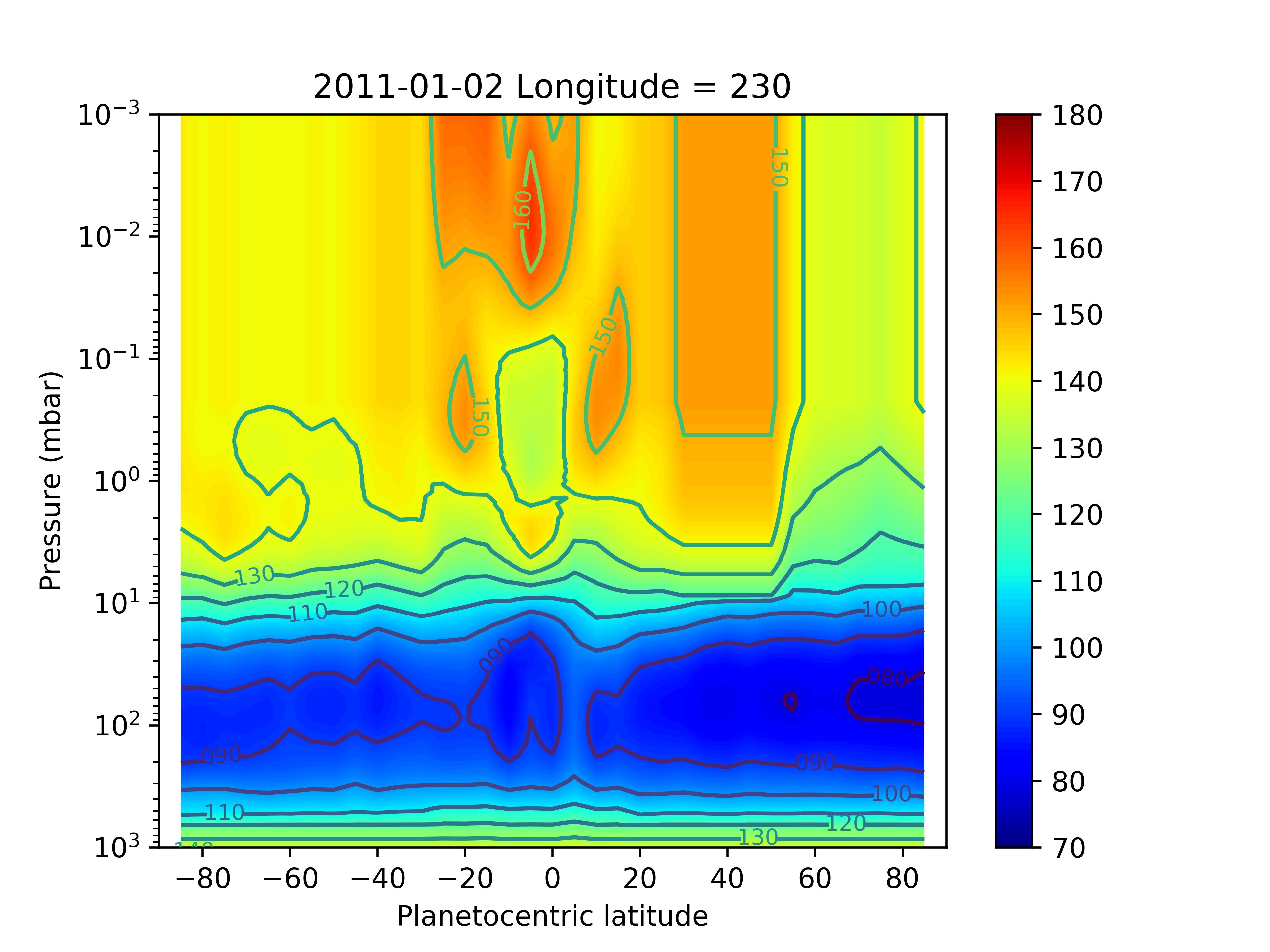}
   \includegraphics[width=9cm,keepaspectratio]{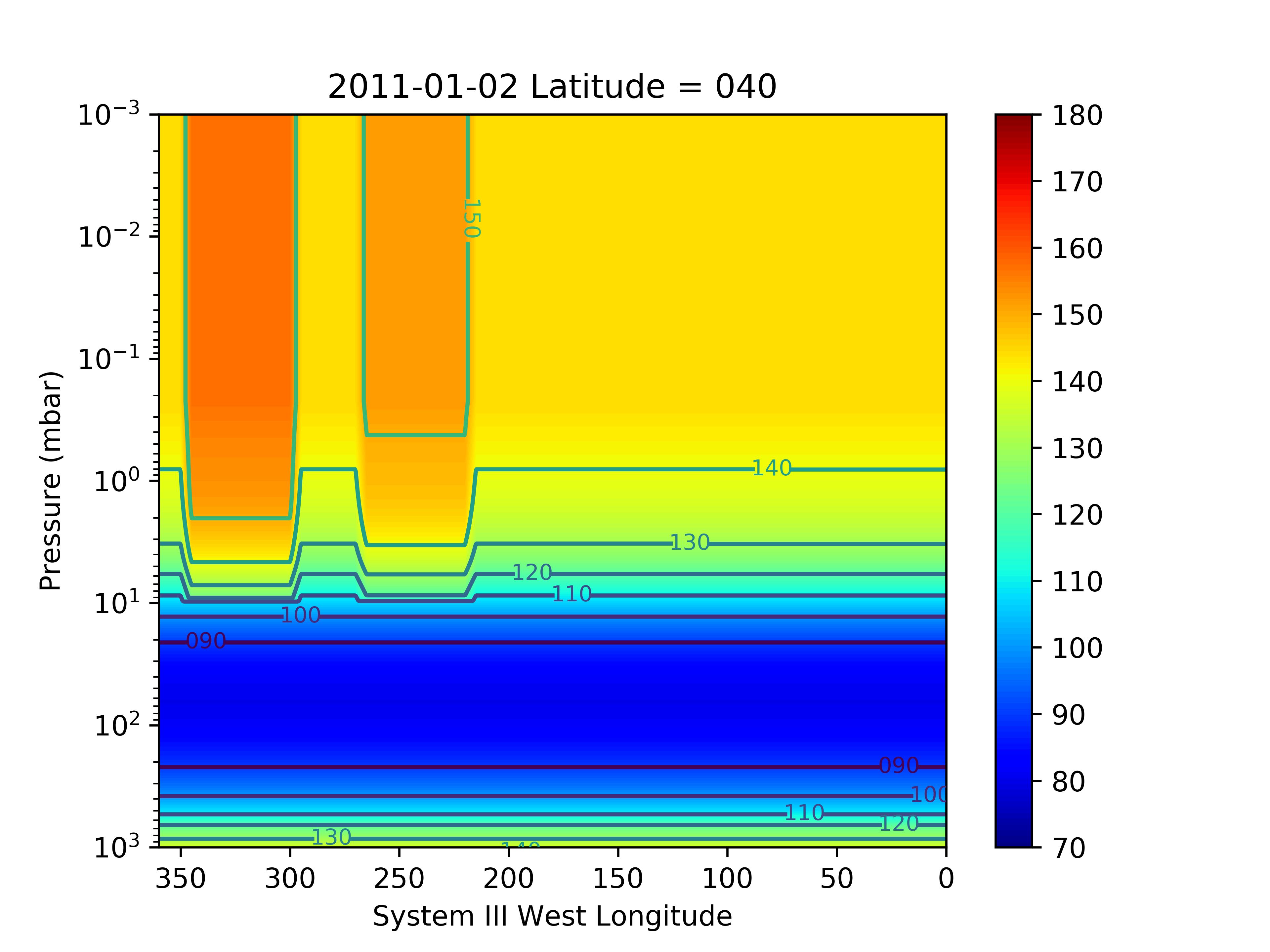}
   \includegraphics[width=9cm,keepaspectratio]{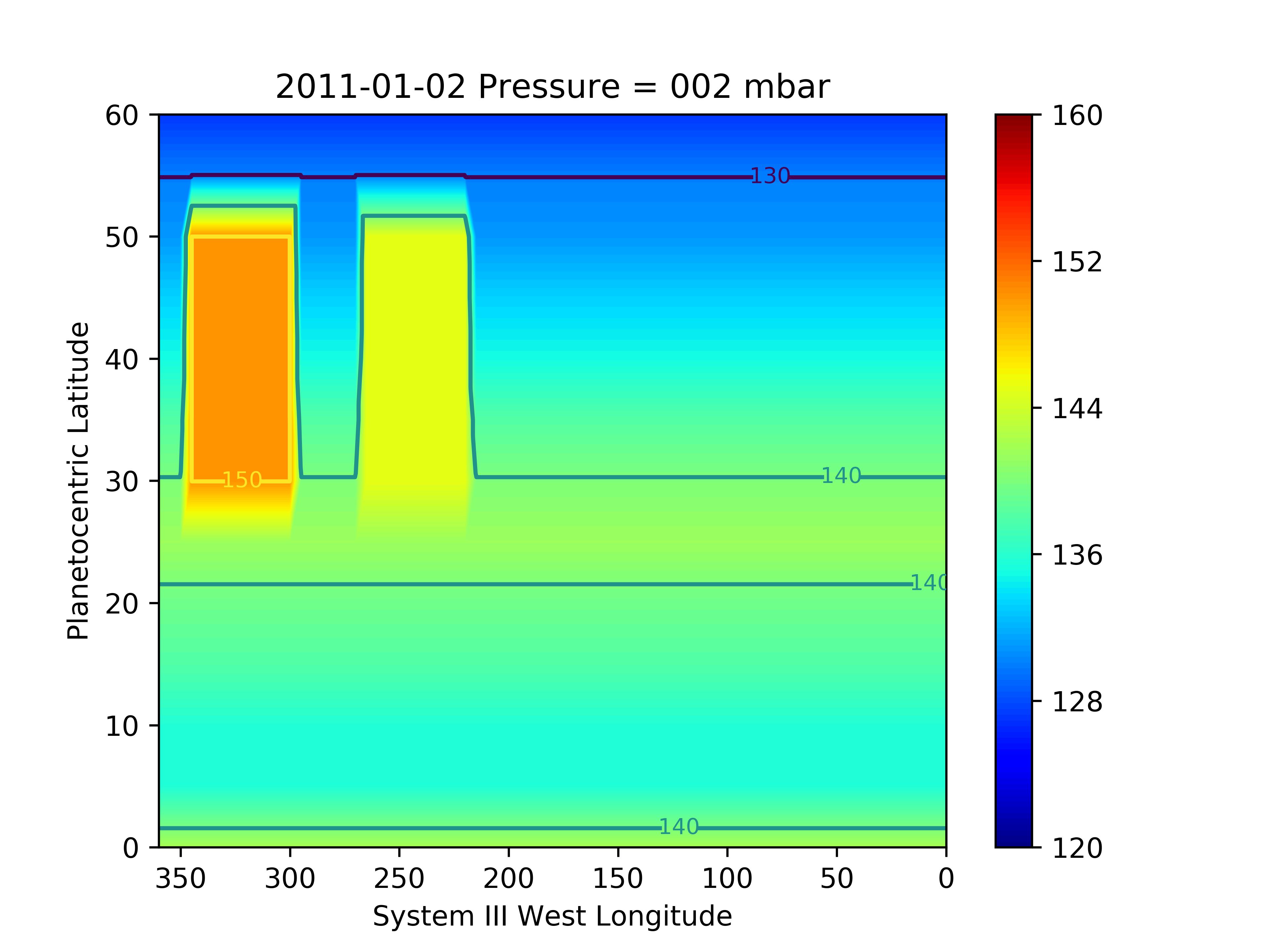}
\end{center}
\caption{3D thermal field used in simulations accounting for 10\,K temperature increases in B1 and 5\,K for B2. B1 is located between 300W-355W, and B2 between 220W-265W (when accounting for longitudinal smearing during the PACS integration), and both between 30\,N and 50\,N \citep{Fletcher2012}. The top panels are meridional cross-sections isolating B1 (left) and B2 (right). The bottom left panel is a zonal cross-section at 40\,N, and the bottom right is a pressure cross-section at 2\,mbar, in which both beacons can be identified.}
\label{Beacons} 
\end{figure*}

\begin{figure}[!h]
\begin{center}
   \includegraphics[width=8cm,keepaspectratio]{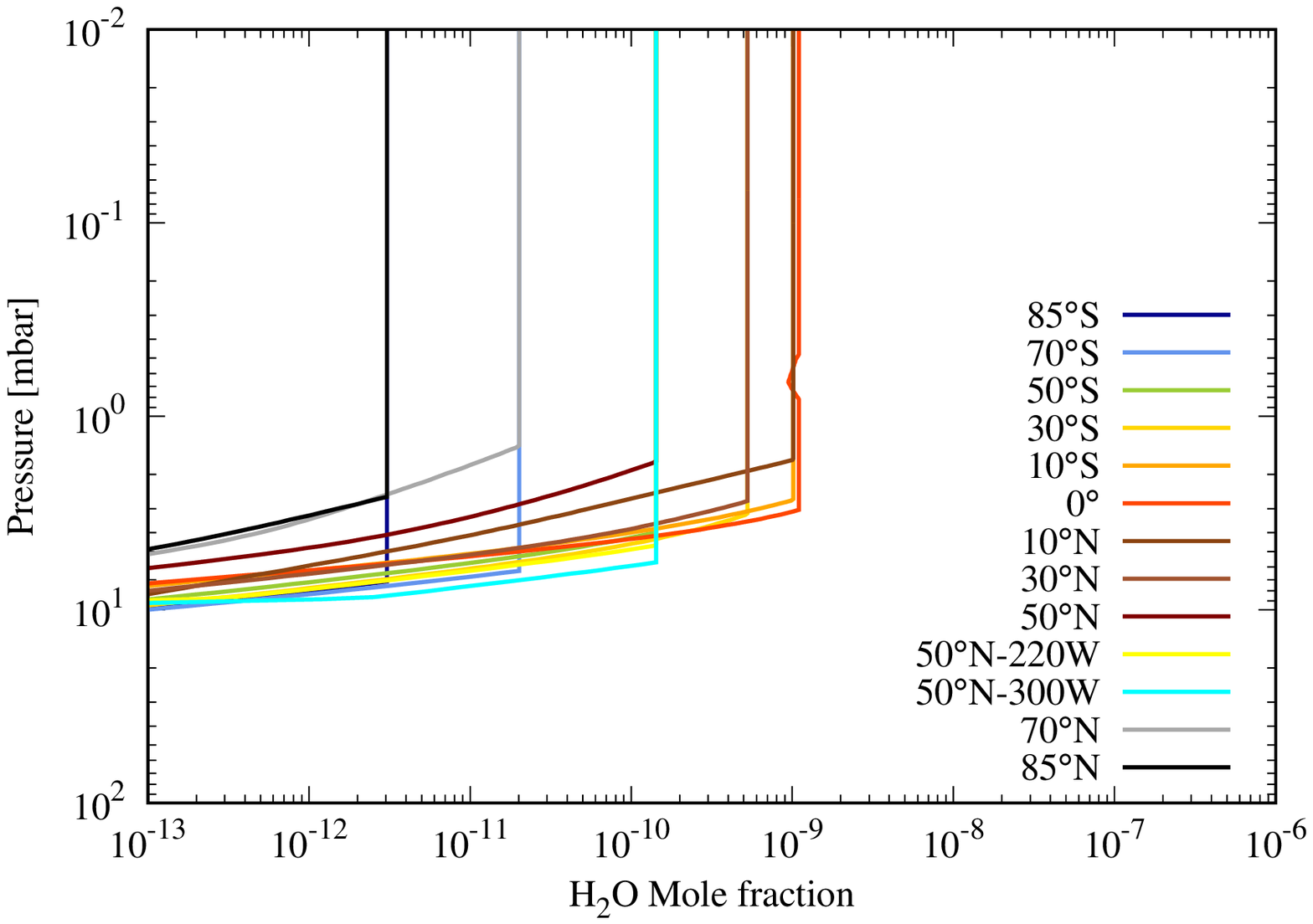}
\end{center}
\caption{H$_2$O vertical profiles of the PACS best fit model (see \fig{PACS_beacon2}), in which the latitudinal distribution is a Gaussian centered around the equator with $\sigma$$=$25\degre, and $y_\mathrm{eq}$$=$1.1\,ppb. The effect of the temperature increases in the B1 and B2 beacons is accounted for when computing the condensation level, as shown by the yellow and cyan lines (corresponding to B1 and B2, respectively). The local equatorial minimum at 1\,mbar results from condensation at a temperature minimum caused by the Saturn quasi-periodic oscillation (see \fig{Thermal_field_prestorm}). The corresponding line area and residual maps are shown in \fig{PACS_beacon2}.
}
\label{beacon_H2O} 
\end{figure}

   According to Cassini/CIRS observations made on the same day as the PACS data were recorded (and two days after the HIFI observations), the B1 and B2 beacons were located in the following longitude ranges: 300W-340W for B1, and 220W-250W for B2. They were thus both in the PACS field-of-view. Each one was located $\sim$halfway between the central meridian (289\degre) and the planetary limb (west for B1 and east for B2). Only B2 was in the HIFI field-of-view, located around the central meridian (227\degre). At that time, CIRS did only observe their northern edges and an increase of $\sim$5-10K had been measured there at the mbar level \citep{Fletcher2012}. Given the limited spatial coverage of the Cassini data (and the absence of ground-based thermal data) at that date, we have assumed a latitudinal extent of B1 and B2 from 30\degre~N to 50\degre~N (planetocentric), and tried several temperature increase combinations within B1 and B2 compared to adjacent longitudes. As the emission excess seen in the data is more obvious in the northwest than in the northeast, we assume a stronger temperature increase in B1 compared to B2. We have tested the following combinations: $+6$\,K/$+3$\,K, $+10$\,K/$+5$\,K, $+15$\,K/$+10$\,K, and $+20$\,K/$+15$\,K for B1/B2 (respectively), and applied these temperature increases uniformly above the 10-mbar level at all latitudes/longitudes in B1/B2, following the vertical trends in temperature found in these early development stages of B1 and B2 (see Fig.~8a in \citealt{Fletcher2012}). The choice of 10\,mbar as a cut-off pressure for the temperature increase was guided by the following argument: this level is below the H$_2$O condensation level, so that we do not probe this level, while it is still well above the level where the continuum is generated. It is therefore a pragmatic way to ensure that the temperature changes in the beacon will impact the H$_2$O lines. For B1 and B2, we also account in the modeling of the thermal field for the longitudinal smearing of 15\degre~caused by the PACS integration time. Our thermal field thus becomes a 3D field in our subsequent modeling (see \fig{Beacons}), and we also recompute the condensation level of H$_2$O in the beacons according to this new thermal field (see \fig{beacon_H2O}). 
   
\begin{figure}[!h]
\begin{center}
   \includegraphics[width=10cm,keepaspectratio]{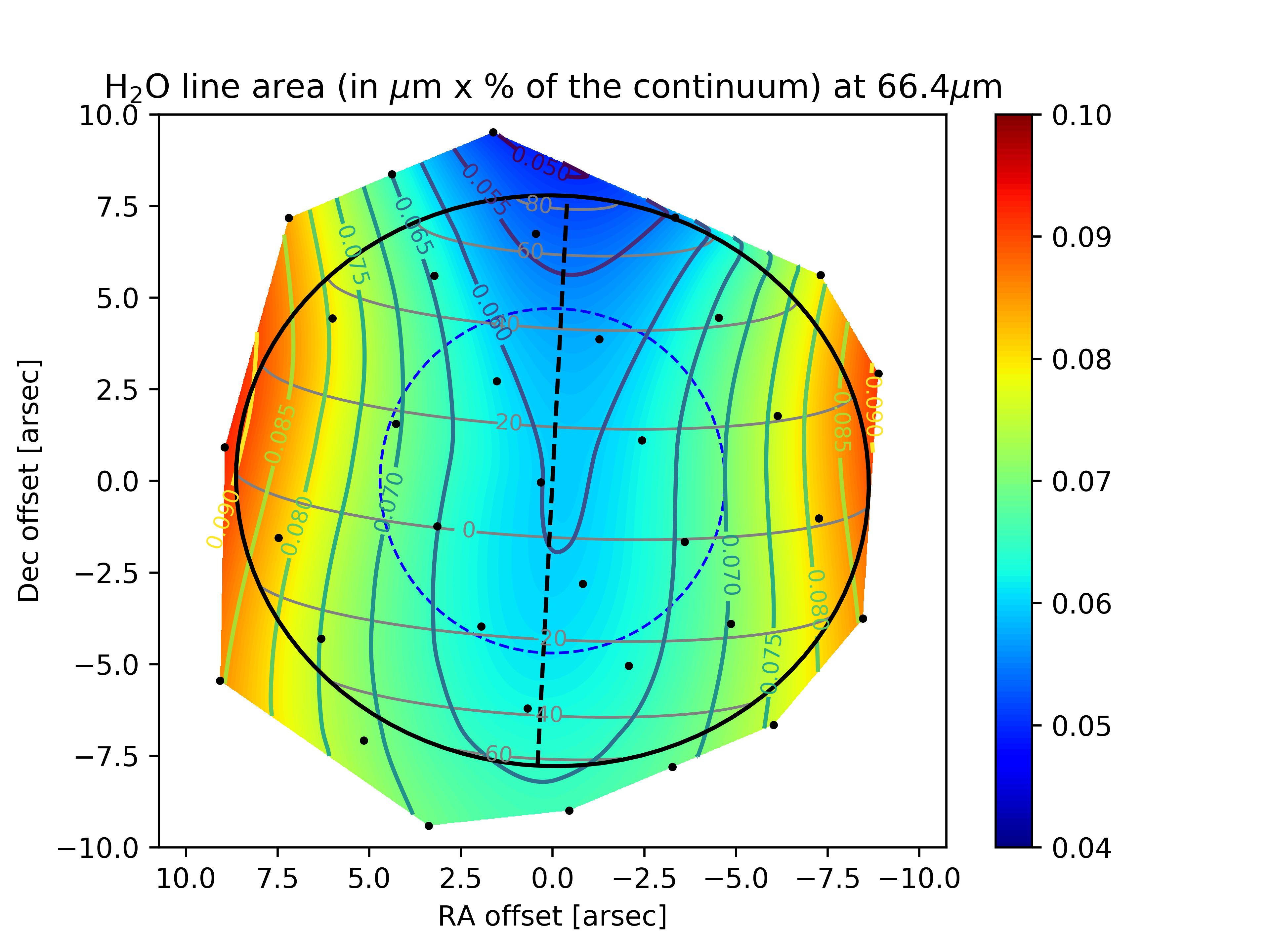}
   \includegraphics[width=10cm,keepaspectratio]{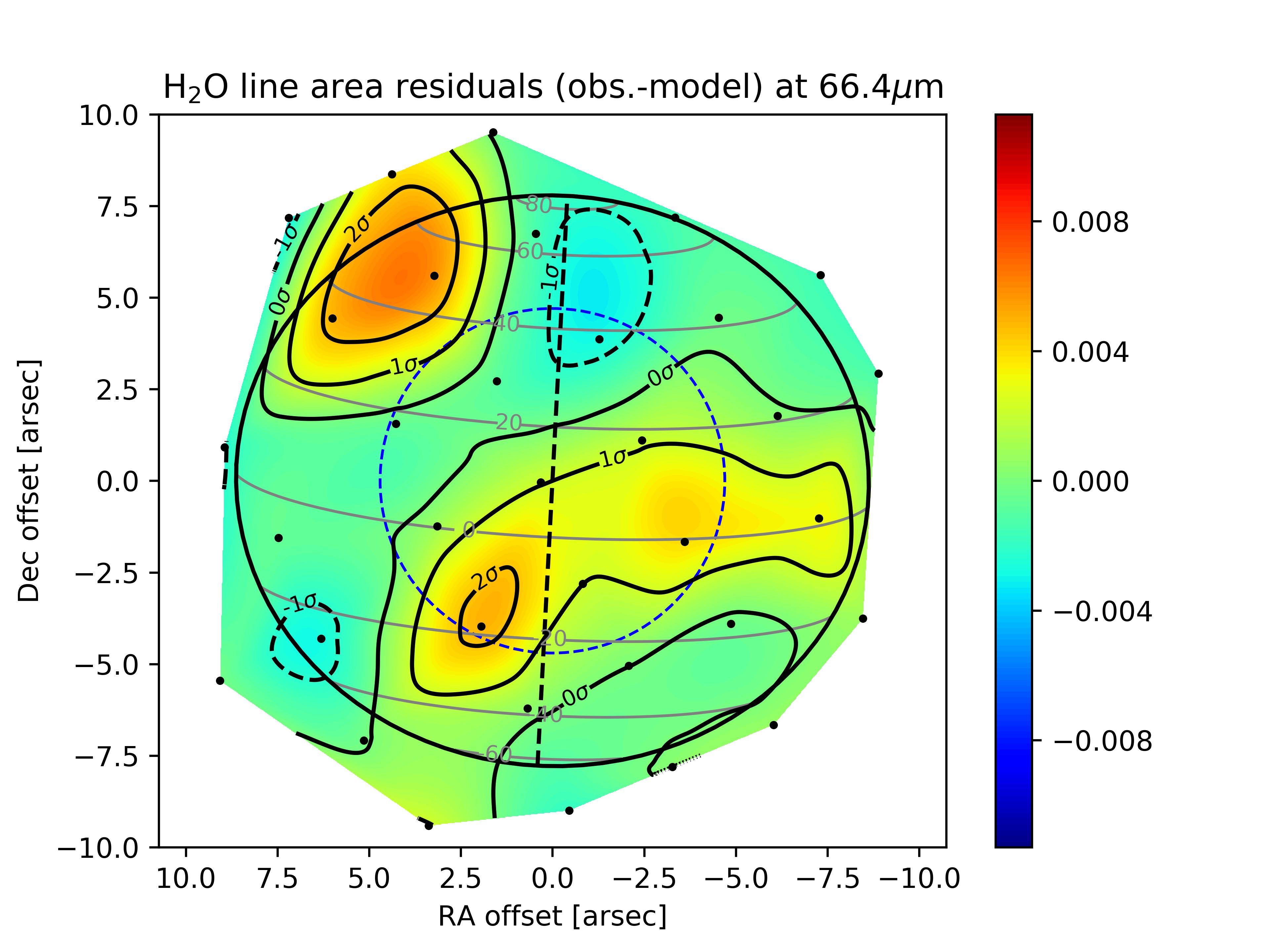}
\end{center}
\caption{Same as \fig{PACS_Uniform} for a Gaussian distribution of H$_2$O around the equator with $y_\mathrm{eq}$$=$1.1\,ppb and $\sigma$$=$25\degre, after accounting for the temperature increases in the B1 and B2 stratospheric beacons caused by Saturn's Great Storm of 2010-2011 at the time of the PACS observations. The overall fit is very good, even if some emission excess remains in the northwest limb. Increasing the temperature further in B2 would improve the fit in this region, but degrade the fit at the North Pole because of the large beam of PACS. 
}
\label{PACS_beacon2} 
\end{figure}

\begin{figure*}[!h]
\begin{center}
   \includegraphics[width=17cm,keepaspectratio]{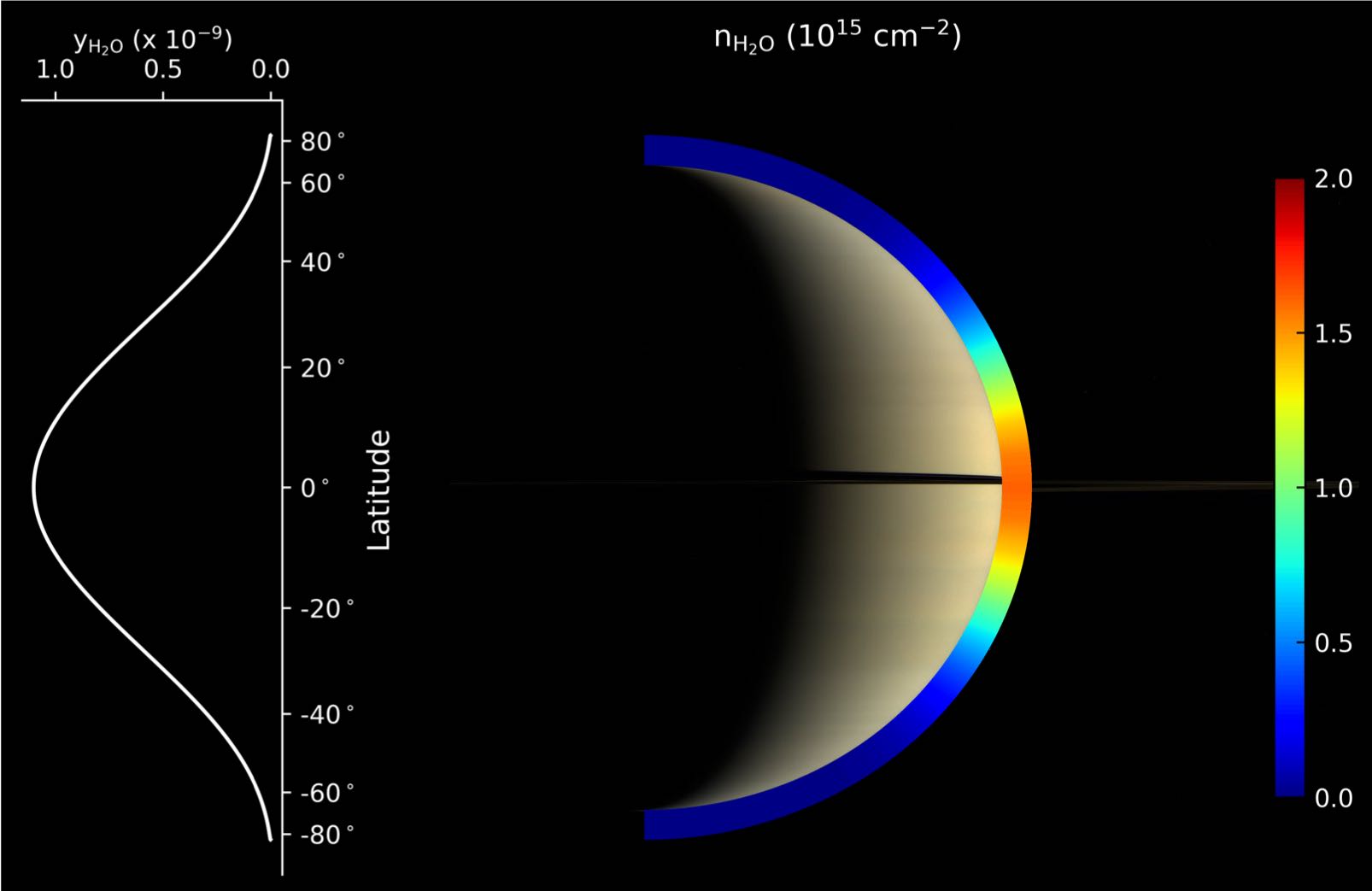}
\end{center}
\caption{Saturn's stratospheric H$_2$O meridional distribution as derived from \textit{Herschel}-PACS mapping observations on January 2, 2011: (left) mole fraction above the local condensation level as a function of latitude, and (right) corresponding column density as a function of latitude. The distribution is a gaussian centered on the equator with $y_\mathrm{eq}$$=$1.1\,ppb and $\sigma$$=$25\degre, corresponding to the PACS map best-fit model (see \fig{PACS_beacon2}). Background image credits: NASA/JPL-Caltech/Space Science Institute.
}
\label{PACS_beacon2_Coldens} 
\end{figure*}

   We obtain the best-fit results for the Gaussian meridional distribution of H$_2$O with uniform temperature increases in B1 and B2 of 10\,K and 5\,K, respectively. The resulting 3D temperature field is shown in several latitudinal cross-sections in \fig{Beacons}. With this field, we significantly improve the fit to the PACS data and find a range of very good-fit ($\sigma$,$y_\mathrm{eq}$) combinations, with a best fit ($\chi^2/N$$=$1.1) for $\sigma$$=$25\degre~and $y_\mathrm{eq}$$=$1.1\,ppb. The line area and residual maps are displayed in \fig{PACS_beacon2} and the column density as a function of latitude is shown in \fig{PACS_beacon2_Coldens}. We stress that we did not change the H$_2$O abundance within the beacons in this work (only the pressure of the condensation level). The results seem to be indicating some enhancement in H$_2$O abundance from the excess of emission in the observations compared to our simulation. In addition, the temperature profile within the beacons was more complex than our idealized model, with a peak at 0.5\,mbar and a drop at lower pressures \citep{Fletcher2012}. This should result in the need for more H$_2$O in the mbar region to compensate for the fainter emission above the temperature peak. We leave this for future analysis of the beacon emissions. 
   
\begin{figure}[!h]
\begin{center}
   \includegraphics[width=10cm,keepaspectratio]{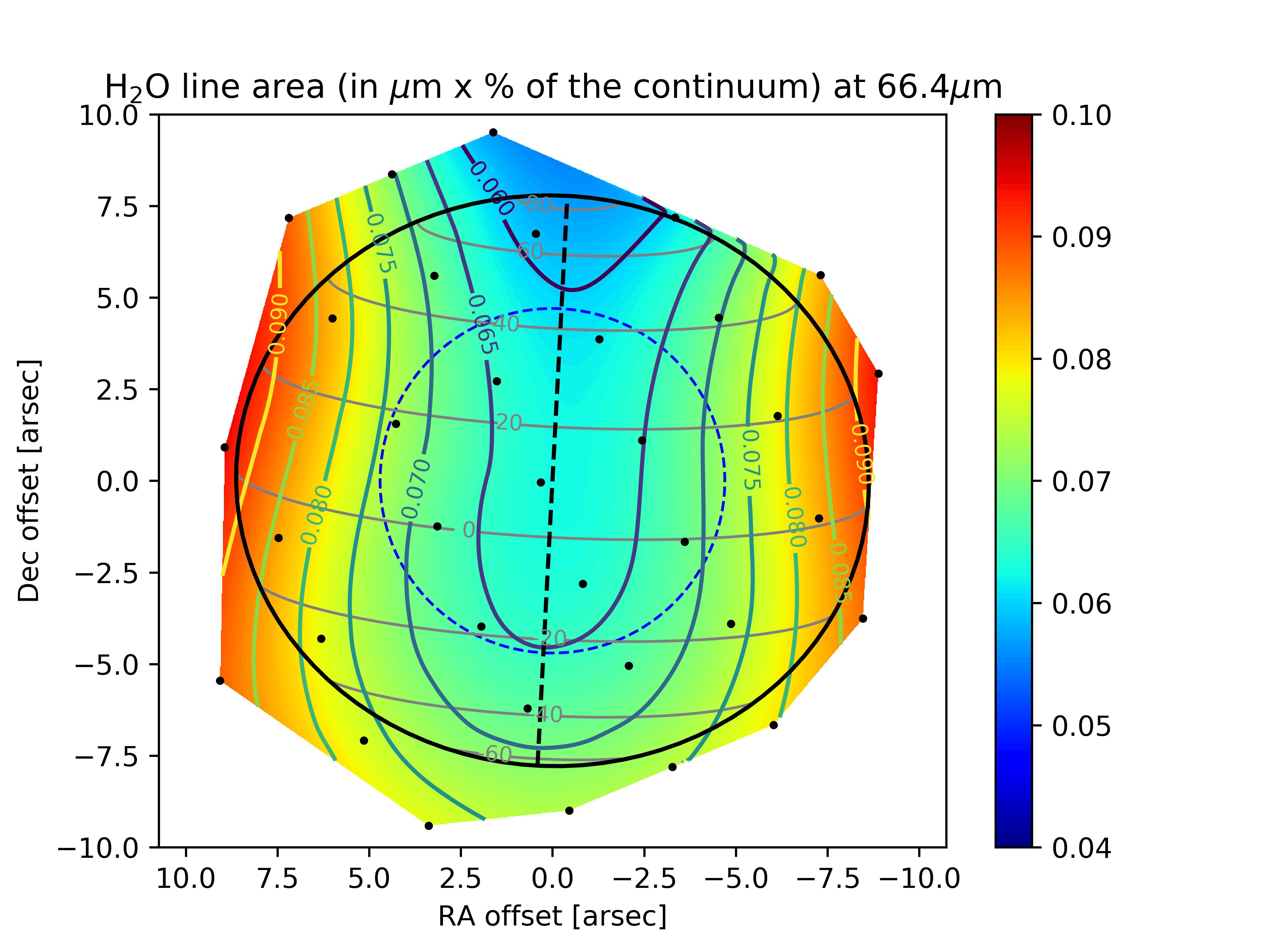}
   \includegraphics[width=10cm,keepaspectratio]{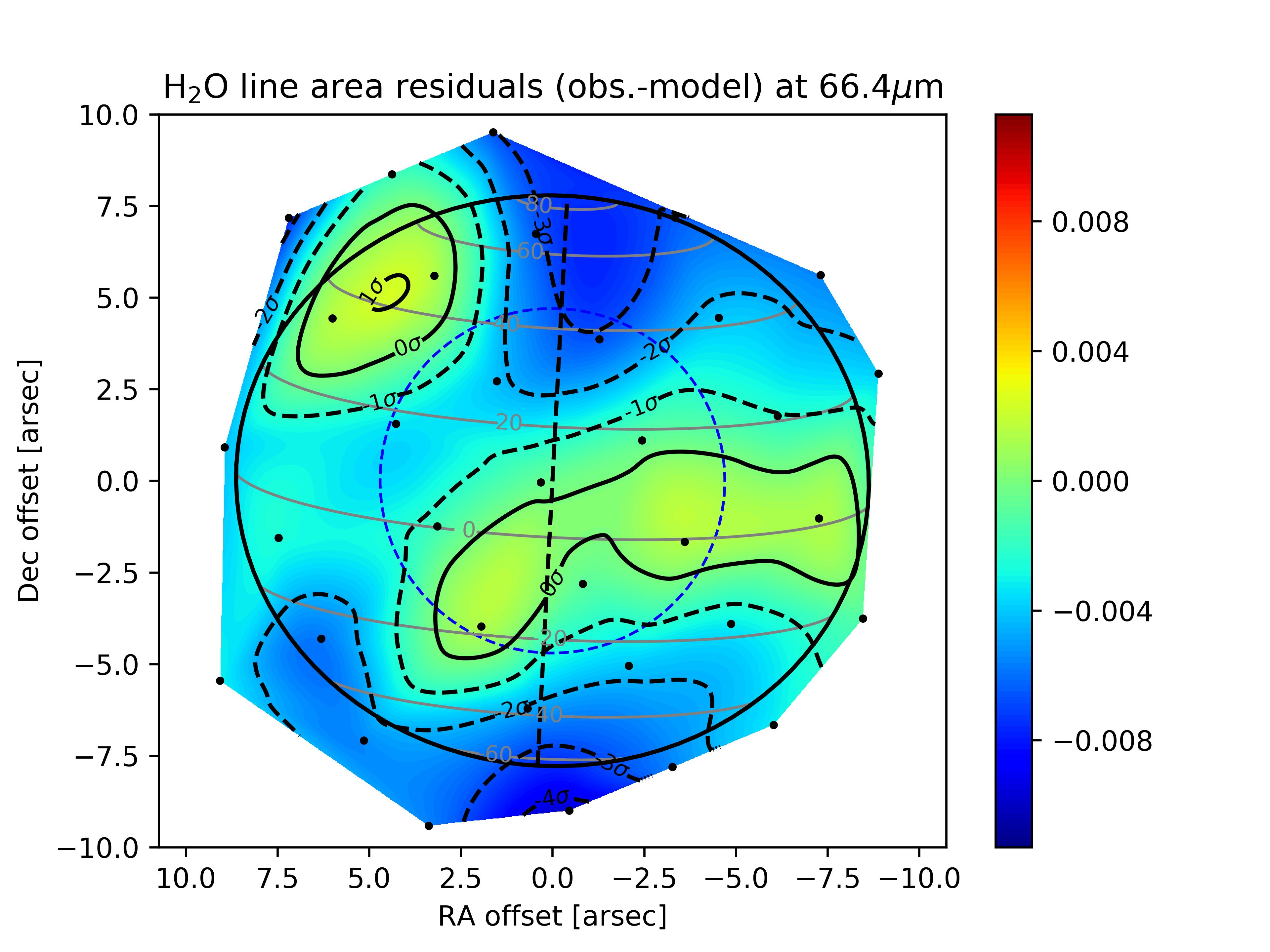}
\end{center}
\caption{Same as \fig{PACS_beacon2}, in which a uniform component is added to the Gaussian distribution of H$_2$O. The model distribution parameters are $y_\mathrm{eq}$$=$1.04\,ppb, $\sigma$$=$25\degre, and $y_\mathrm{min}$$=$0.06\,ppb. The overall fit is degraded compared to \fig{PACS_beacon2}, because of the uniform component of the distribution. 
}
\label{PACS_beacon2_ymin} 
\end{figure}

   With this model, we can even constrain a background and meridionally uniform flux (e.g., IDP) represented by a variable $y_\mathrm{min}$ that can be added to equation~\ref{H2O_gaussian}. We find that it cannot exceed 0.06\,ppb, at the 2-$\sigma$ limit (\fig{PACS_beacon2_ymin}), which is an order of magnitude lower than the disk-averaged contribution of the gaussian distribution. This confirms the small contribution from IDP, as predicted by \citet{Moses2017}. New photochemical simulations are now required to determine the corresponding IDP H$_2$O flux upper limit. 
   
   Accounting for the temperature increase in B2, which is in the HIFI field-of-view does not improve the situation regarding the PACS/HIFI incompatibility, i.e., $\sim$5 times more H$_2$O than in the PACS best-fit model is needed to fit the HIFI line. It remains unchanged in terms of $y_\mathrm{eq}$ for a given $\sigma$, compared to Section~\ref{gaussian_distribution}, and the fit to the HIFI data keeps the same flaws (wings too broad and line center too weak).
   
\begin{figure}[!h]
\begin{center}
   \includegraphics[width=10cm,keepaspectratio]{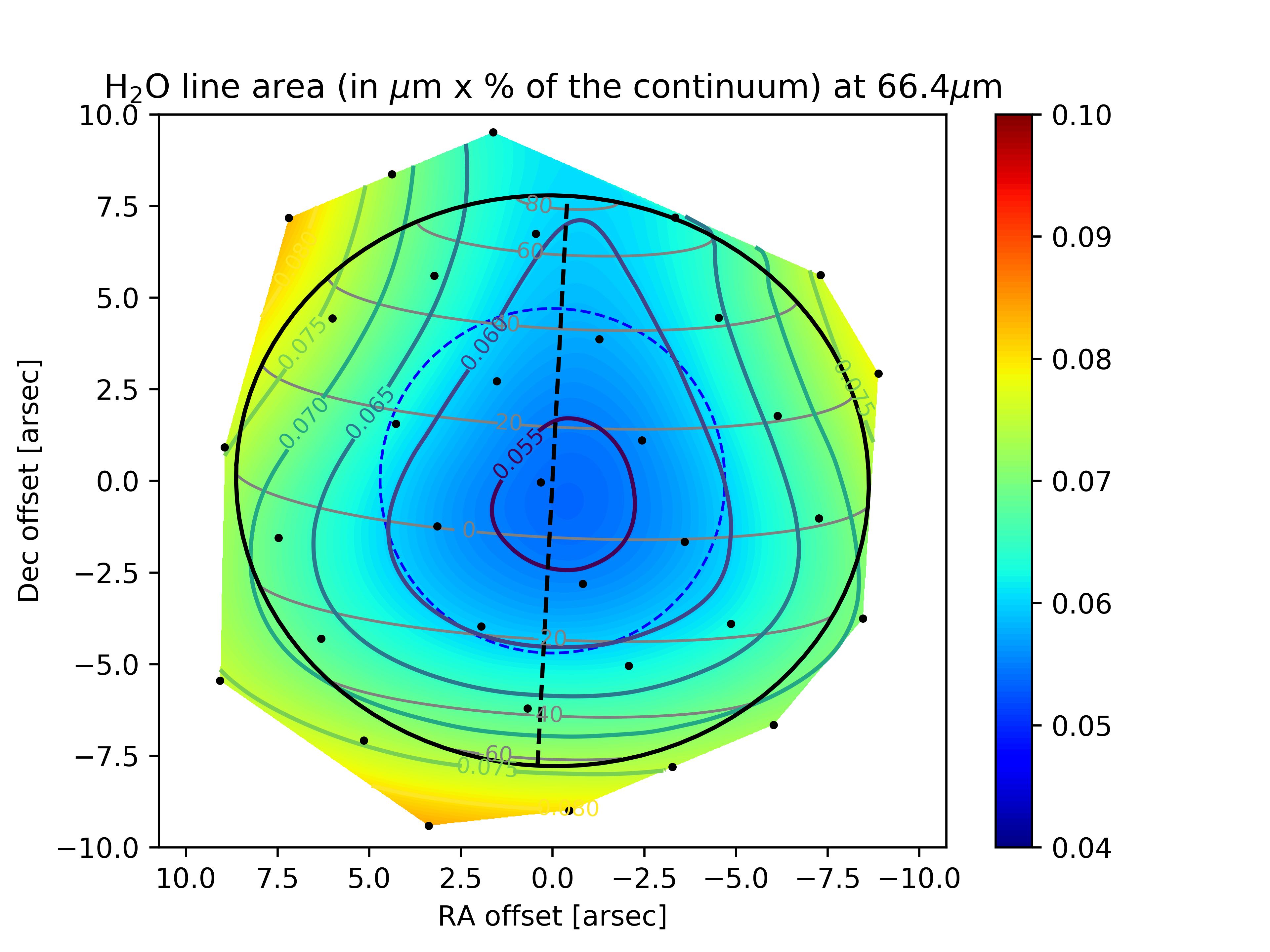}
   \includegraphics[width=10cm,keepaspectratio]{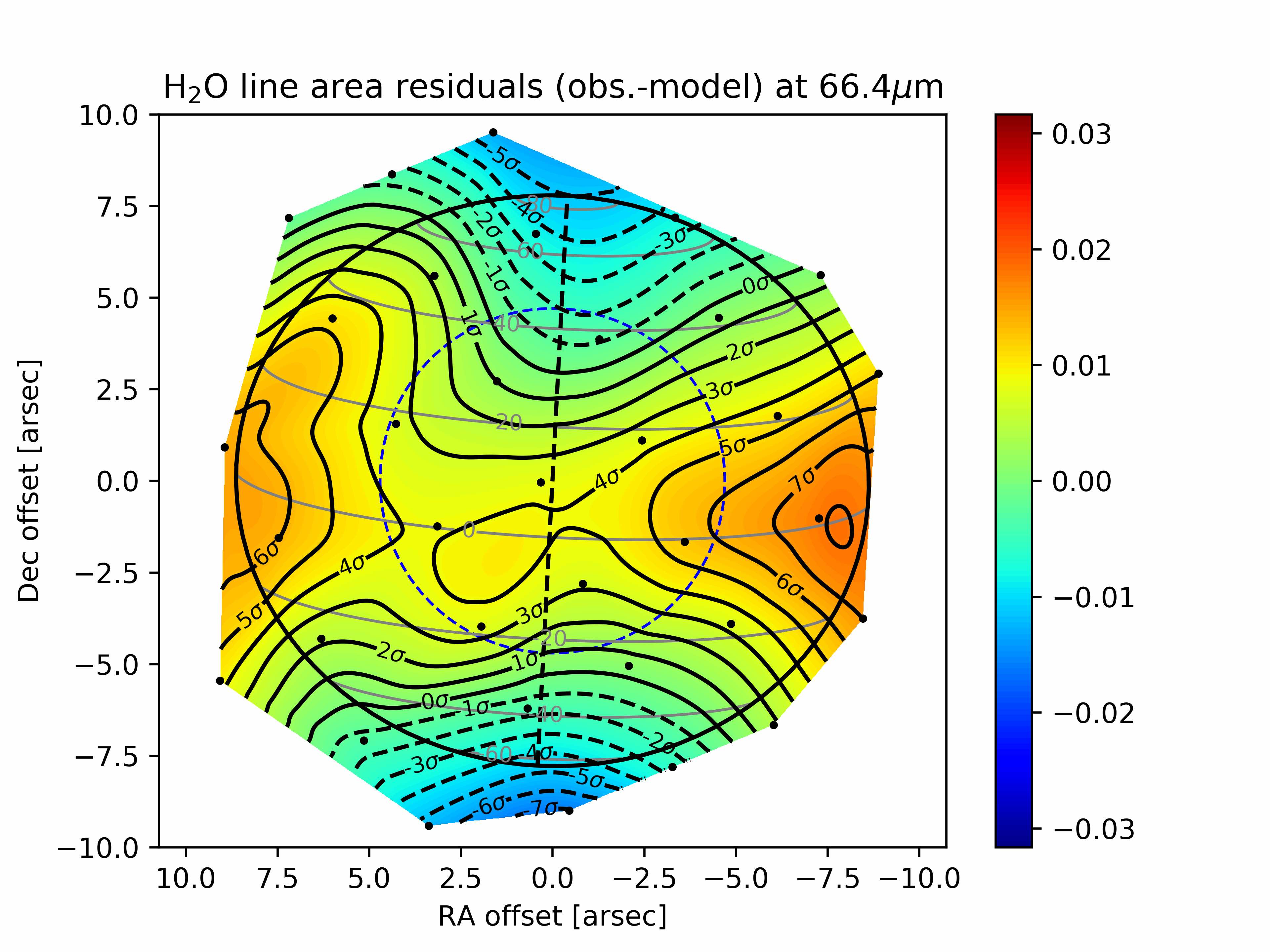}
\end{center}
\caption{Same as \fig{PACS_Uniform} for a meridionally uniform distribution of H$_2$O with a mole fraction set to 4\dix{-10}, after accounting for the temperature increases in the B1 and B2 stratospheric beacons caused by Saturn's Great Storm of 2010-2011 at the time of the PACS observations. The overall fit is still poor ($\chi^2/N$$=$16.2) such that accounting for the temperature increases caused by Saturn's Great Storm of 2010-2011 does not reconcile such an H$_2$O distribution with the data. 
}
\label{Uniform_beacon2} 
\end{figure}

   Now that we have a good fit to the PACS data, we have to briefly turn back to the meridionally uniform distribution model to check whether the improvement in the temperature field induces any improvement in the fit to the data for such a thermal model. \fig{Uniform_beacon2} shows the line area and residual map obtained for an H$_2$O mole fraction of 4\dix{-10} above the condensation level. With a $\chi^2/N$$=$16.2, such a model remains invalid. After leveling out the thermal effects from the line emission map, the uniform distribution still results in high latitudes being too bright and low latitudes being too faint compared to the observations.

   Accounting for the effect on the thermal field of Saturn's Great Storm of 2010-2011 has enabled us to significantly improve the quality of the fit to the PACS data with the Gaussian meridional H$_2$O distribution and confirmed the invalidity of the meridionally uniform distribution. 

   \subsection{Introducing a gradient in the H$_2$O vertical profile above its condensation layer \label{sec_gradient}}   
   In this section, we seek to reconcile the PACS and HIFI observations while building on the improvements in the PACS interpretation that resulted from including Saturn's Great Storm effect in our modeling, by introducing a gradient in the H$_2$O vertical distribution above the condensation level.
      
\begin{figure}[!h]
\begin{center}
   \includegraphics[width=8cm,keepaspectratio]{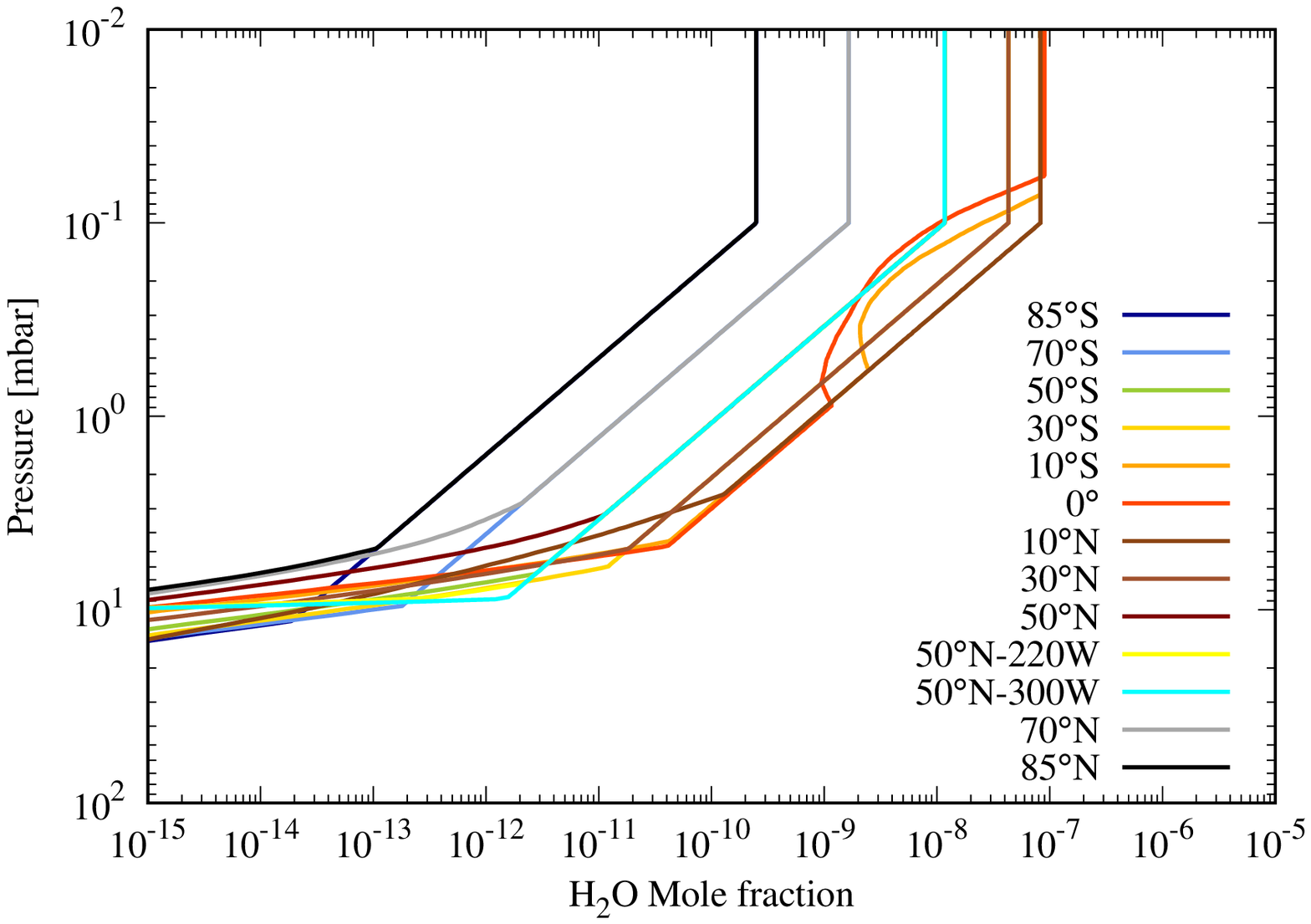}
\end{center}
\caption{H$_2$O vertical profiles of the PACS fit model of \fig{Gradient}, in which the latitudinal distribution is a Gaussian centered around the equator with $\sigma$$=$25\degre, and $y_\mathrm{eq}$$=$9\dix{-8}. The effect of the temperature increases in the B1 and B2 beacons is accounted for when computing the condensation level, and a positive gradient is added for pressures higher than $p_\mathrm{gradient}$$=$0.1\,mbar and down to the condensation level, with a slope $n$$=$$\log(y)/\log(p)$$=$2. The value of $y_\mathrm{eq}$ applies to $p$$<$$p_\mathrm{gradient}$. Condensation between 0.1 and 1\,mbar occurs in the equatorial region as a result of a local temperature minimum caused by the quasi-periodic oscillation (see \fig{Thermal_field_prestorm}). The corresponding line area and residual maps are shown in \fig{Gradient}. 
}
\label{abundance_gradient7} 
\end{figure}

   As the HIFI line probes slightly higher than the PACS line, and as the HIFI line is too broad in the wings and too faint at line center, we introduce a parametrized vertical gradient in the H$_2$O profile at all latitudes to make it less abundant above the condensation level and more abundant at higher altitude. This way, the empirical H$_2$O vertical profiles will be closer to physical profiles computed with photochemical models (e.g., \citealt{Ollivier2000,Moses2000}). As the PACS data require less H$_2$O than the HIFI data, we see this gradient as a means to reconcile the two datasets. To achieve this, we introduce two additional parameters: a cut-off pressure $p_\mathrm{gradient}$, above which the H$_2$O abundance is held constant, and $n$ which is the $\log(y)/\log(p)$ slope of the profile between the cut-off pressure and the condensation level, as already used in previous works \citep{Marten2005,Rezac2014}. We test several combinations of $p_\mathrm{gradient}$ and $n$, with values of $p_\mathrm{gradient}$ ranging from 0.01\,mbar to 1\,mbar ($p_\mathrm{gradient}$ is always smaller than the pressure of the condensation level, whatever the latitude) and values of $n$ ranging from 0.5 to 3 (an example is shown in \fig{abundance_gradient7}). 
   
\begin{figure}[!h]
\begin{center}
   \includegraphics[width=10cm,keepaspectratio]{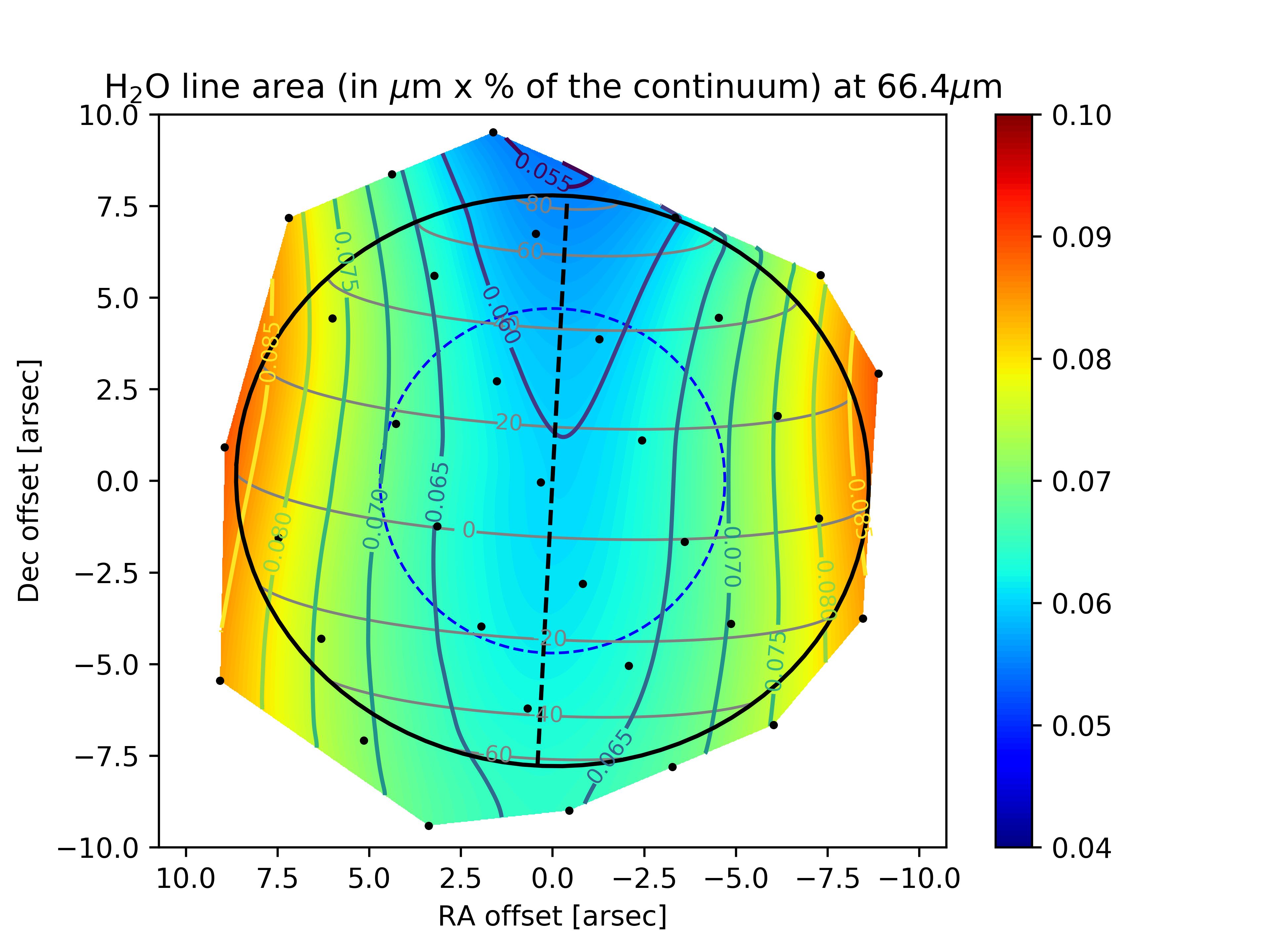}
   \includegraphics[width=10cm,keepaspectratio]{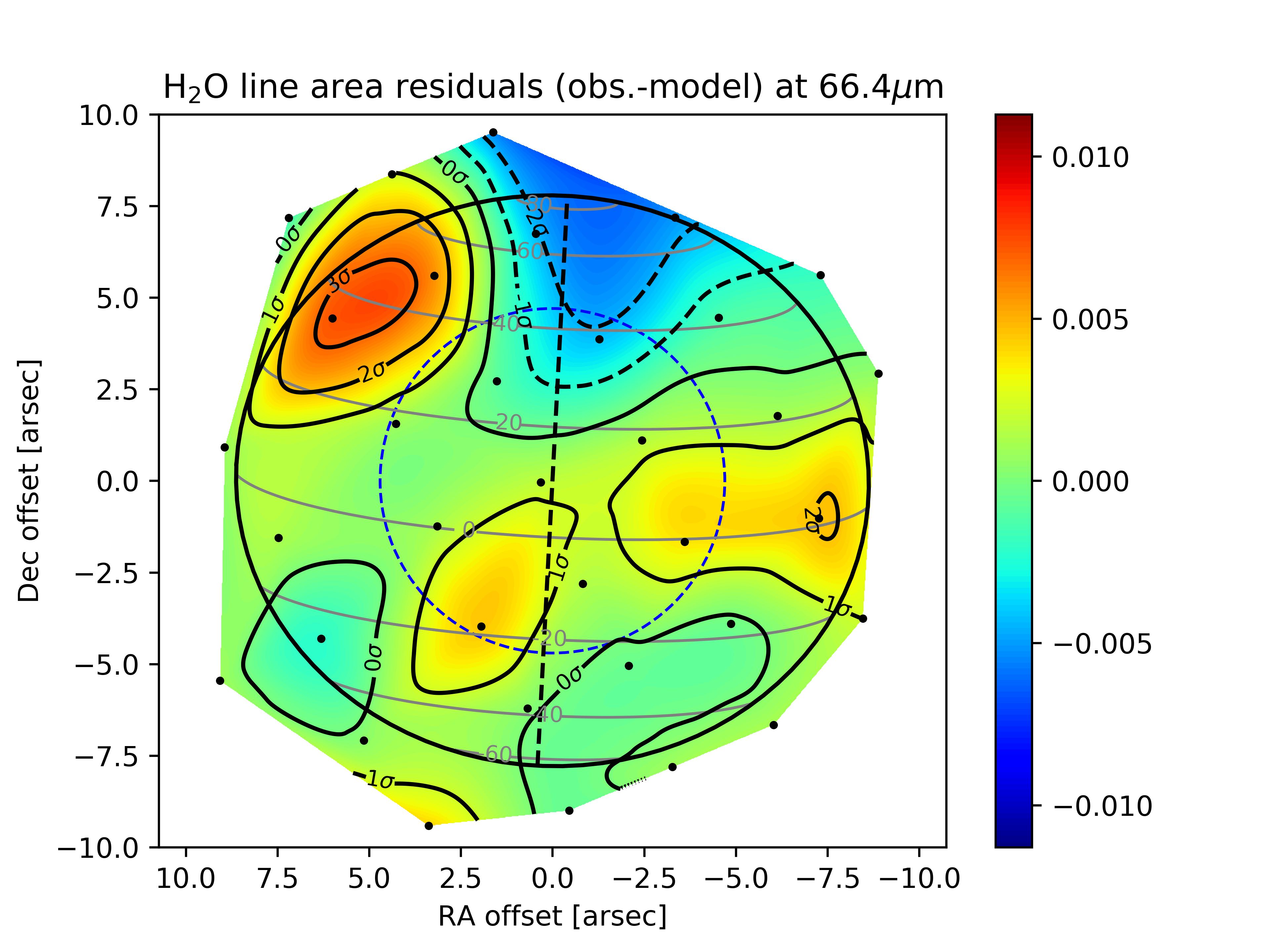}
\end{center}
\caption{Same as \fig{PACS_Uniform} for a Gaussian distribution of H$_2$O around the equator with $y_\mathrm{eq}$$=$9\dix{-8} and $\sigma$$=$25\degre, after accounting for the temperature increases in the B1 and B2 stratospheric beacons caused by Saturn's Great Storm of 2010-2011 at the time of the PACS observations, and a positive gradient in the H$_2$O profile above the condensation level (with $n$$=$2 and $p_\mathrm{gradient}$$=$0.1\,mbar). More details about the vertical profiles can be found in \fig{abundance_gradient7}. The overall fit remains good compared to \fig{PACS_beacon2}, and enables reducing to a factor of $\sim$2 the inconsistency between the $y_\mathrm{eq}$ values derived from PACS and HIFI. 
}
\label{Gradient} 
\end{figure}

\begin{figure}[!h]
\begin{center}
   \includegraphics[width=9cm,keepaspectratio]{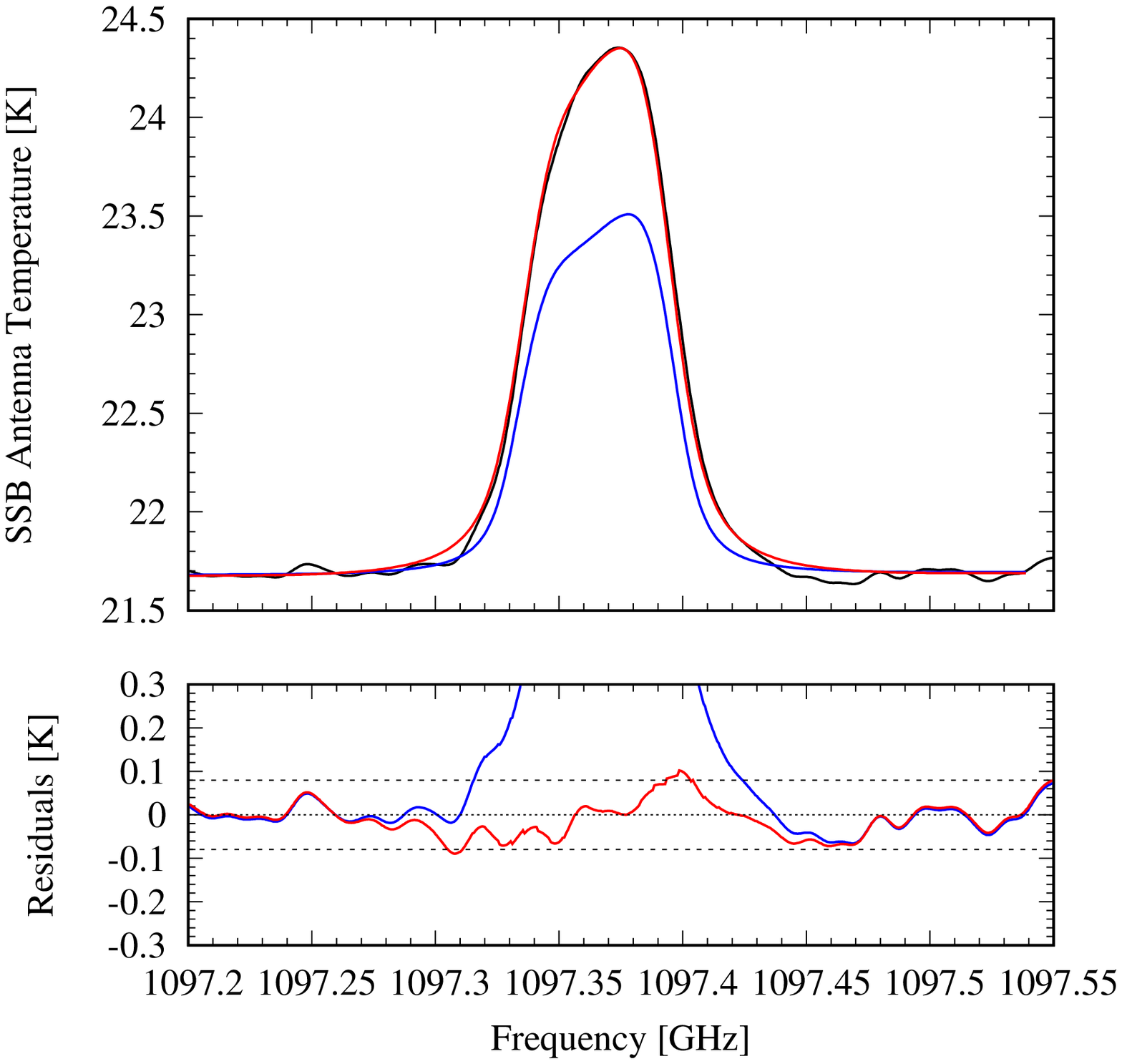}
\end{center}
\caption{Best fit to the HIFI data for a Gaussian distribution of H$_2$O around the equator with $y_\mathrm{eq}$$=$2.2\dix{-7} and $\sigma$$=$25\degre, after accounting for the effect of the temperature increases in B1 and B2 and a positive gradient for pressures higher than $p_\mathrm{gradient}$$=$0.1\,mbar and down to the condensation level with a slope $n$$=$$\log(y)/\log(p)$$=$2. The best-fit model to the PACS data from \fig{Gradient} is shown for comparison. Same layout as in \fig{Contribution} left.
}
\label{HIFI_gradient7} 
\end{figure}

   We find that adding a positive gradient in the H$_2$O vertical profile above the condensation layer reduces the incompatibility between the HIFI and PACS $y_\mathrm{eq}$ values from a factor of $\sim$5 down to a factor of $\sim$2.4 for $n$$=$2--3 and $p_\mathrm{gradient}$$<$0.3\,mbar in the range of values we have tested. An example for PACS is shown in \fig{Gradient} for $n$$=$2, $p_\mathrm{gradient}$$=$0.1\,mbar, $y_\mathrm{eq}$$=$9\dix{-8} and $\sigma$$=$25\degre. The corresponding vertical profiles as a function of latitude are displayed in \fig{abundance_gradient7}. The corresponding best fit of the HIFI line is obtained for the same set of parameters, except $y_\mathrm{eq}$$=$2.2\dix{-7} (see \fig{HIFI_gradient7}).

   \subsection{Final remarks on the HIFI/PACS discrepancy \label{discrepancy}}   
   As it stands, we find no H$_2$O distribution that enables us to fully reconcile the HIFI and PACS data in terms of H$_2$O abundance (through the value of $y_\mathrm{eq}$), even if introducing a positive gradient above the condensation level significantly improves the situation.  
   
   There are some systematic and random errors we have not considered that may explain at least in part the remaining discrepancy between the HIFI and PACS derived abundances. For instance, there is some scatter in the PH$_3$ disk-averaged abundances derived from observations (e.g., \citealt{Weisstein1996} vs. \citealt{Fletcher2012}), its meridional distribution was only measured once by \citet{Fletcher2009b} and any temporal evolution is therefore unconstrained. As noted in Section~\ref{RT_model}, the PH$_3$ distribution influences the continuum of the 1097\,GHz line (but not of the 66.4\,$\mu$m one), and thus the derivation of the H$_2$O abundance from this line, quite significantly. The thermal field also bears some uncertainties ($\sim$3\,K at the tropopause and $\sim$4\,K at 1-5\,mbar, as detailed in \citet{Fletcher2018b}). Given that we implement a full 3D thermal field and that the opacities of the observed H$_2$O lines are of the same order but not strictly identical, the uncertainties on the thermal field may be an additional cause of discrepancy between the HIFI and PACS lines.
   
   Finally, we stress that the meridional distribution may be more complicated than the one given by Eq.~\ref{H2O_gaussian}, and the vertical gradient may also vary with latitude. This study can thus be seen as a stepping stone for future computations of physical profiles with more complete altitude-latitude photochemical models.

\section{Discussion \label{Discussion}}
In situ measurements carried out by several Cassini instruments during the end-of-mission orbits and final plunge into the atmosphere (all in 2017) have provided strong evidence for material infalling from the rings onto Saturn's atmosphere, either in gaseous or solid form, either neutral or ionized \citep{Hsu2018,Mitchell2018,Waite2018}. While small infalling icy grains, constrained to $\pm$2\degre~around the equatorial plane, rain onto Saturn with a modest mass flux of 5\,kg/s \citep{Mitchell2018}, H$_2$O gas seems to be feeding Saturn's upper atmosphere at a much higher rate. From Cassini's Ion and Neutral Mass Spectrometer (INMS), \citet{Waite2018} found that a global integrated mass flux of 4800--45000\,kg/s of gaseous material was raining onto the planet's atmosphere at the time of the measurements, with 24$\pm$5\% H$_2$O, within a latitudinal band of 8\degre~centered on the equator. Using data from the same instrument, \citet{Perry2018} find an even larger influx of 2--20\dix{4}\,kg/s. This is 2 to 4 orders of magnitude higher than the influx of H$_2$O of 7--25\,kg/s derived from previous disk-averaged observations and models \citep{Feuchtgruber1997,Bergin2000,Moses2005,Hartogh2011}. 
Therefore, this material infalling from the inner ring system, seen in 2017 by Cassini, cannot be the source of the H$_2$O observed by \textit{Herschel} in December 2010 and January 2011. 

An earlier Cassini finding provided another credible candidate: Enceladus and its H$_2$O plumes \citep{Hansen2006,Porco2006,Waite2006}. 
Cassini's first observations of the plumes at the south pole of the moon venting out significant amounts of H$_2$O  triggered a modeling effort to evaluate the fate of the outgassed molecules \citep{Jurac2007,Cassidy2010,Moore2010}. For instance, \citet{Cassidy2010} found that neutral/neutral scattering was the main cause of spreading of the molecules in Saturn's system and predicted that Enceladus plumes formed an H$_2$O neutral torus at the orbital distance of the satellite. This torus was confirmed by the observations of \citet{Hartogh2011}. According to \citet{Cassidy2010}, this torus is likely to be a significant source of H$_2$O for Saturn's rings, the planet itself and even Titan. The latter was confirmed by \citet{Moreno2012}. At Saturn, the meridional distribution of infalling neutral H$_2$O originating from Enceladus plumes is predicted by \citet{Cassidy2010} to be Gaussian and centered on the planet equator with a HWHM $\sigma$$\sim$15\degre. The disk-averaged mass flux was estimated to 6\dix{5}\Fluxunit, i.e., $\sim$8\,kg.s$^{-1}$, by \citet{Hartogh2011} by fitting the Enceladus source to the absorption caused by the Enceladus torus on the Saturn H$_2$O line at 557\,GHz. Comparing this 6\dix{5}\Fluxunit~flux to the required external flux invoked to explain Saturn's water \Result{1}{0.5}{6} \citep{Moses2000}, \citet{Hartogh2011} concluded that the Enceladus plumes were the likely source of Saturn's stratospheric H$_2$O. 

In this paper, we have analyzed a \textit{Herschel}-PACS map of H$_2$O emission coming from Saturn's stratosphere. We have presented clear evidence that H$_2$O in Saturn's stratosphere is not meridionally uniform, and that a distribution in which its abundance peaks at the equator and decreases exponentially towards higher latitudes fits the data. This result enables us to identify Enceladus as the main source of Saturn's stratospheric H$_2$O, and to reject the IDP as a primary source. However, a faint meridionally uniform background flux cannot be ruled out by our observations. Our data indicate that this background is an order of magnitude lower than the disk-averaged contribution of the Enceladus source, in agreement with the prediction from \citet{Moses2017}. 

The difference we find between the predicted width of the influx meridional distribution with the observed one allow us to compute a rough estimate of meridional eddy mixing at low-to-mid latitudes. If we assume that $K_{yy}\sim L^2/t$, we then take $L$$\sim$10500\,km as the distance between 15\degre~latitude and 25\degre~latitude. These latitudes correspond to the HWHMs of the input flux from \citet{Cassidy2010} and our best model. For the diffusion timescale between those latitudes, we take $t$ equal to the downward diffusion timescale from the top of the atmosphere, where the material is delivered, to the condensation level (150\,y according to \citealt{Moses2017}), where H$_2$O was mapped by \textit{Herschel}. We find that $K_{yy}$$\sim$$2$\dix{8}\Kunit, and this value is nominally lower by a factor of 10 than early modeling results from \citet{Friedson2011}. A more accurate derivation of the input fluxes as a function of latitude at Saturn that are responsible for the observations will require additional simulation work with 2D photochemical models \citep{Hue2015,Hue2016,Hue2018}.

\section{Conclusion \label{Conclusion}}
The findings of our paper can be summarized as follow:
\begin{itemize}
  \item We have mapped Saturn's stratospheric H$_2$O emission with \textit{Herschel}-PACS on January 2, 2011, and obtained a disk-averaged spectrum of the H$_2$O line at 1097\,GHz with \textit{Herschel}-HIFI on December 31, 2010.
  \item A meridionally uniform distribution of H$_2$O above its condensation level does not fit the data, which implies that IDP are not the primary source of Saturn's stratospheric H$_2$O.
  \item When accounting for the 3D temperature field at the time of our observations, including the effect of Saturn's Great Storm of 2010-2011, we find that the data are reasonably well reproduced with a meridional distribution of H$_2$O that is Gaussian-shaped and centered around the equator. The best fit is obtained for an equatorial mole fraction of 1.1\,ppb and a HWHM of 25\degre. This type of distribution points to the Enceladus plumes as the primary source of Saturn's observed stratospheric H$_2$O.
  \item We can place an upper limit on a meridionally uniform background source like IDP about 10 times fainter than the main equatorial source, in agreement with theoretical predictions by \citet{Moses2017}. 
  \item We can improve the compatibility between the results obtained from the HIFI and PACS observations by adding a positive gradient above the condensation level in the H$_2$O vertical profiles, in an attempt to bring our profiles closer to physical profiles. However, we do not manage to fully reconcile the disk-resolved PACS data and the disk-averaged HIFI data with our empirical models, probably because of remaining systematic and random errors and because of the simplicity of our meridional distribution model. Full 2D photochemical modeling is now necessary to move to the next step.
\end{itemize}

It remains to be seen what meridional distribution of input fluxes are required to reproduce the \textit{Herschel} observations, and 2D photochemical modeling is required at this stage, starting for example with those of \citet{Hue2015,Hue2016,Hue2018}. Given their extraordinary sensitivity, bandwidth, and spectral and spatial resolutions, future complementary ALMA and \textit{JWST} observations will certainly help to shed light on the recent extraordinary influx of exogenic material from the inner ring system to Saturn's stratosphere, as captured by Cassini in its final orbits \citep{Waite2018,Perry2018,Mitchell2018,Hsu2018}. These future observations \citep{Lellouch2008,Norwood2016a}, as well as possibly direct in situ measurements \citep{Mousis2014,Mousis2016,Mousis2018}, will also help to improve our understanding of the outstanding and more general question of the origin of exogenic species in giant planet atmospheres.


\section*{Acknowledgements}
This work was supported by the Programme National de Plan\'etologie (PNP) of CNRS/INSU, co-funded by CNES. T. Cavali\'e was supported by a CNES fellowship at the beginning of this work. Support for G. Orton was provided by NASA through an award issued by the Jet Propulsion Laboratory, California Institute of Technology.
  
PACS has been developed by a consortium of institutes led by MPE (Germany) and including UVIE (Austria); KU Leuven, CSL, IMEC (Belgium); CEA, LAM (France); MPIA (Germany); INAF-IFSI/OAA/OAP/OAT, LENS, SISSA (Italy); IAC (Spain). This development has been supported by the funding agencies BMVIT (Austria), ESA-PRODEX (Belgium), CEA/CNES (France), DLR (Germany), ASI/INAF (Italy), and CICYT/MCYT (Spain).
  
HIFI has been designed and built by a consortium of institutes and university departments from across Europe, Canada and the United States under the leadership of SRON Netherlands Institute for Space Research, Groningen, The Netherlands and with major contributions from Germany, France and the US. Consortium members are: Canada: CSA, U.Waterloo; France: CESR, LAB, LERMA, IRAM; Germany: KOSMA, MPIfR, MPS; Ireland, NUI Maynooth; Italy: ASI, IFSI-INAF, Osservatorio Astrofisico di Arcetri-INAF; Netherlands: SRON, TUD; Poland: CAMK, CBK; Spain: Observatorio Astron\'omico Nacional (IGN), Centro de Astrobiolog\'ia (CSIC-INTA). Sweden: Chalmers University of Technology - MC2, RSS \& GARD; Onsala Space Observatory; Swedish National Space Board, Stockholm University - Stockholm Observatory; Switzerland: ETH Zurich, FHNW; USA: Caltech, JPL, NHSC.

The \textit{Herschel} spacecraft was designed, built, tested, and launched under a contract to ESA managed by the \textit{Herschel}/\textit{Planck} Project team by an industrial consortium under the overall responsibility of the prime contractor Thales Alenia Space (Cannes), and including Astrium (Friedrichshafen) responsible for the payload module and for system testing at spacecraft level, Thales Alenia Space (Turin) responsible for the service module, and Astrium (Toulouse) responsible for the telescope, with in excess of a hundred subcontractors.

HCSS / HSpot / HIPE is a joint development (are joint developments) by the \textit{Herschel} Science Ground Segment Consortium, consisting of ESA, the NASA \textit{Herschel} Science Center, and the HIFI, PACS and SPIRE consortia.


\bibliographystyle{aa} 

\end{document}